\documentclass{pasj01}
\Received{$\langle$reception date$\rangle$}
\Accepted{$\langle$acception date$\rangle$}
\Published{$\langle$publication date$\rangle$}

\begin{document}

\title{SILVERRUSH. II. First Catalogs and Properties of\\
$\sim 2,000$ Ly$\alpha$ Emitters and Blobs at $z\sim 6-7$\\
Identified over the $14-21$ deg$^2$ Sky} 
\author{Takatoshi Shibuya\altaffilmark{1}, 
Masami Ouchi\altaffilmark{1,2}, 
Akira Konno\altaffilmark{1,3}, 
Ryo Higuchi\altaffilmark{1,4}, 
Yuichi Harikane\altaffilmark{1,4}, 
Yoshiaki Ono\altaffilmark{1}, 
Kazuhiro Shimasaku\altaffilmark{3,5}, 
Yoshiaki Taniguchi\altaffilmark{6}, 
Masakazu A. R. Kobayashi\altaffilmark{7}, 
Masaru Kajisawa\altaffilmark{8}, 
Tohru Nagao\altaffilmark{8}, 
Hisanori Furusawa\altaffilmark{9}, 
Tomotsugu Goto\altaffilmark{10}, 
Nobunari Kashikawa\altaffilmark{9,11}, 
Yutaka Komiyama\altaffilmark{9,11}
Haruka Kusakabe\altaffilmark{3}, 
Chien-Hsiu Lee\altaffilmark{12}, 
Rieko Momose\altaffilmark{10}, 
Kimihiko Nakajima\altaffilmark{13}, 
Masayuki Tanaka\altaffilmark{9,11}, 
Shiang-Yu Wang\altaffilmark{14}, 
and 
Suraphong Yuma\altaffilmark{15}
}

\altaffiltext{1}{Institute for Cosmic Ray Research, The University of Tokyo, 5-1-5 Kashiwanoha, Kashiwa, Chiba 277-8582, Japan }
\altaffiltext{2}{Kavli Institute for the Physics and Mathematics of the Universe (Kavli IPMU, WPI), University of Tokyo, Kashiwa, Chiba 277-8583, Japan}
\altaffiltext{3}{Department of Astronomy, Graduate School of Science, The University of Tokyo, 7-3-1 Hongo, Bunkyo, Tokyo 113-0033, Japan} 
\altaffiltext{4}{Department of Physics, Graduate School of Science, The University of Tokyo, 7-3-1 Hongo, Bunkyo, Tokyo 113-0033, Japan} 
\altaffiltext{5}{Research Center for the Early Universe, Graduate School of Science, The University of Tokyo, 7-3-1 Hongo, Bunkyo, Tokyo 113-0033, Japan}
\altaffiltext{6}{The Open University of Japan, Wakaba 2-11, Mihama-ku, Chiba 261-8586, Japan}
\altaffiltext{7}{Faculty of Natural Sciences, National Institute of Technology, Kure College, 2-2-11 Agaminami, Kure, Hiroshima 737-8506, Japan}
\altaffiltext{8}{Research Center for Space and Cosmic Evolution, Ehime University, Bunkyo-cho 2-5, Matsuyama 790-8577, Japan}
\altaffiltext{9}{National Astronomical Observatory, Mitaka, Tokyo 181-8588, Japan}
\altaffiltext{10}{Institute of Astronomy, National Tsing Hua University, 101 Section 2, Kuang-Fu Road, Hsinchu 30013, Taiwan} 
\altaffiltext{11}{The Graduate University for Advanced Studies (SOKENDAI), 2-21-1 Osawa, Mitaka, Tokyo 181-8588}
\altaffiltext{12}{Subaru Telescope, NAOJ, 650 N Aohoku Pl., Hilo, HI 96720, USA}
\altaffiltext{13}{European Southern Observatory, Karl-Schwarzschild-Str. 2, D-85748, Garching bei Munchen, Germany}
\altaffiltext{14}{Academia Sinica, Institute of Astronomy and Astrophysics, 11F of AS/NTU Astronomy-Mathematics Building, No.1, Sec. 4, Roosevelt Rd, Taipei 10617, Taiwan}
\altaffiltext{15}{Department of Physics, Faculty of Science, Mahidol University, Bangkok 10400, Thailand }
\altaffiltext{\ddag}{Based on data obtained with the Subaru Telescope. The Subaru Telescope is operated by the National Astronomical Observatory of Japan.}
\email{shibyatk@icrr.u-tokyo.ac.jp}

\KeyWords{early universe --- galaxies: formation --- galaxies: high-redshift}

\maketitle

\begin{abstract}
We present an unprecedentedly large catalog consisting of 2,230 $\gtrsim L^*$ Ly$\alpha$ emitters (LAEs) at $z=5.7$ and $6.6$ on the $13.8$ and $21.2$ deg$^2$ sky, respectively, that are identified by the SILVERRUSH program with the first narrowband imaging data of the Hyper Suprime-Cam (HSC) survey. We confirm that the LAE catalog is reliable on the basis of 96 LAEs whose spectroscopic redshifts are already determined by this program and the previous studies. This catalogue is also available on-line. Based on this catalogue, we derive the rest-frame Ly$\alpha$ equivalent-width distributions of LAEs at $z\simeq5.7-6.6$ that are reasonably explained by the exponential profiles with the scale lengths of $\simeq120-170$\,\AA, showing no significant evolution from $z\simeq5.7$ to $z\simeq6.6$. We find that $275$ LAEs with a large equivalent width (LEW) of $>240$ \AA\ are candidates of young-metal poor galaxies and AGNs. We also find that the fraction of LEW LAEs to all ones is $4$\% and $21$\% at $z\simeq5.7$ and $z\simeq6.6$, respectively. Our LAE catalog includes 11 Ly$\alpha$ blobs (LABs) that are LAEs with spatially extended Ly$\alpha$ emission whose profile is clearly distinguished from those of stellar objects at the $\gtrsim 3\sigma$ level. The number density of the LABs at $z=6-7$ is $\sim 10^{-7}-10^{-6}$ Mpc$^{-3}$, being $\sim 10-100$ times lower than those claimed for LABs at $z\simeq 2-3$, suggestive of disappearing LABs at $z\gtrsim 6$, albeit with the different selection methods and criteria for the low and high-$z$ LABs.
\end{abstract}

\section{Introduction}\label{sec_intro}

Ly$\alpha$ Emitters (LAEs) are one of important populations of high-$z$ star-forming galaxies in the paradigm of the galaxy formation and evolution. Such galaxies are thought to be typically young (an order of $100$ Myr; e.g., \cite{2007ApJ...660.1023F, 2007ApJ...671..278G, 2007ApJ...660.1023F}), compact (an effective radius of $<1$ kpc; e.g., \cite{2009ApJ...701..915T, 2012ApJ...753...95B}), less-massive (a stellar mass of $10^8-10^9 M_\odot$; e.g., \cite{2010MNRAS.402.1580O,2011ApJ...733..114G}), metal-poor ($\simeq0.1$ of the solar metallicity; e.g., \cite{2012ApJ...745...12N, 2013ApJ...769....3N, 2014MNRAS.442..900N, 2016arXiv160503436K}), less-dusty than Lyman break galaxies (e.g., \cite{2011ApJ...736...31B, 2015ApJ...800L..29K}), and a possible progenitor of Milky Way mass galaxies (e.g., \cite{2011ApJ...740...71D}). In addition, LAEs are used to probe the cosmic reionizaiton, because ionizing photons escaped from a large number of massive stars formed in LAEs contribute to the ionization of intergalactic medium (IGM; e.g., \cite{2001ApJ...563L...5R, 2006ApJ...647L..95M, 2006PASJ...58..313S, 2006ApJ...648....7K,2008ApJS..176..301O, 2010ApJ...723..869O, 2010ApJ...711..928C,2010ApJ...725..394H, 2011ApJ...734..119K,2012ApJ...752..114S, 2014ApJ...797...16K, 2015MNRAS.451..400M, 2015MNRAS.451..400M, 2017arXiv170302501O, 2017arXiv170302985Z}). 

LAEs have been surveyed by imaging observations with dedicated narrow-band (NB) filters for a prominent redshifted Ly$\alpha$ emission (e.g., \cite{2002ApJ...576L..25A, 2004ApJ...617L...5M, 2003PASJ...55L..17K, 2005PASJ...57..165T, 2007ApJ...667...79G, 2011ApJ...740L..31E, 2012ApJ...744..110C}). In large LAE sample constructed by the NB observations, two rare Ly$\alpha$-emitting populations have been identified: large equivalent width (LEW) LAEs, and spatially extended Ly$\alpha$ LAEs, Ly$\alpha$ blobs (LABs). 

LEW LAEs are objects with a large Ly$\alpha$ equivalent width (EW) of $\gtrsim240$\,\AA\, which are not reproduced with the normal \authorcite{1955ApJ...121..161S} stellar initial mass function (e.g., \cite{2002ApJ...565L..71M}). Such an LEW is expected to be originated from complicated physical processes such as (i) photoionization by young and/or low-metallicity star-formation, (ii) photoionization by active galactic nucleus (AGN), (iii) photoionization by external UV sources (QSO fluorescence), (iv) collisional excitation due to strong outflows (shock heating), (v) collisional excitation due to gas inflows (gravitational cooling), and (vi) clumpy ISM (see e.g., \cite{2017MNRAS.465.1543H}). The highly-complex radiative transfer of Ly$\alpha$ in the interstellar medium (ISM) makes it difficult to understand the Ly$\alpha$ emitting mechanism (\cite{1991ApJ...370L..85N, 2006MNRAS.367..979H, 2008ApJ...678..655F,2013ApJ...766..124L,2009ApJ...704.1640L,2007ApJ...657L..69L,2010ApJ...716..574Z,2012arXiv1209.5842Y,2013arXiv1302.7042D,2013arXiv1308.1405Z}). 

LABs are spatially extended Ly$\alpha$ gaseous nebulae in the high-$z$ Universe (e.g., \cite{2000ApJ...532..170S, 2004AJ....128..569M,2009ApJ...702..554P, 2009MNRAS.400L..66M,2011MNRAS.410L..13M, 2012ApJ...748..125P,2012ApJ...752...86P,2013ApJ...762...38P,2014Natur.506...63C,2015ApJ...804...26A,2015Sci...348..779H,2015ApJ...799...62P, 2015ApJ...809..163A,2017ApJ...837...71C}). The origins of LABs (LAEs with a diameter $\simeq20-400$ kpc) are also explained by several mechanisms: (1) resonant scattering of Ly$\alpha$ photons emitted from central sources in dense and extended neutral hydrogen clouds (e.g., \cite{2011Natur.476..304H}), (2) cooling radiation from gravitationally heated gas in collapsed halos (e.g., \cite{2000ApJ...534...11H}), (3) shock heating by galactic superwind originated from starbursts and/or AGN activity (e.g., \cite{2000ApJ...532L..13T}), (4) galaxy major mergers (e.g., \cite{2013ApJ...773..151Y}), and (5) photoionization by external UV sources (QSO fluorescence; e.g., \cite{2005ApJ...628...61C}). Moreover, LABs have been often discovered in over-density regions at $z\simeq2-3$ (e.g., \cite{2009ApJ...693.1579Y,2010ApJ...719.1654Y,2011MNRAS.410L..13M}). Thus, such LABs could be closely related to the galaxy environments, and might be liked to the formation mechanisms of central massive galaxies in galaxy protoclusters. 

During the last decades, Suprime-Cam (SCam) on the Subaru telescope has led the world on identifying such rare Ly$\alpha$-emitting populations at $z\gtrsim6$ (LEW LAEs; e.g., \cite{2008ApJ...680..100N, 2012ApJ...761...85K}; LABs; e.g., \cite{2009ApJ...696.1164O,2015ApJ...808..139S}). However, the formation mechanisms of these rare Ly$\alpha$-emitting populations are still controversial due to the small statistics. While LEW LAEs and LABs at $z\simeq2-5$ have been studied intensively with a sample of $\gtrsim100$ sources, only a few sources have been found so far at $z\gtrsim6$. Large-area NB data are required to carry out a statistical study on LEW LAEs and LABs at $z\gtrsim6$. 

In March 2014, the Subaru telescope has started a large-area NB survey using a new wide field of view (FoV) camera, Hyper Suprime-Cam (HSC) in a Subaru strategic program (SSP; \cite{2017arXiv170405858A}). In the five-year project, HSC equipped with four NB filters of ${\it NB}387$, ${\it NB}816$, ${\it NB}921$, and ${\it NB}101$ will survey for LAEs at $z\simeq2.2$, 5.7, 6.6, and 7.3, respectively. The HSC SSP NB survey data consist of two layers; Ultradeep (UD), and Deep (D), covering 2 fields (UD-COSMOS, UD-SXDS), and 4 fields (D-COSMOS, D-SXDS, D-DEEP2-3, D-ELAIS-N1), respectively. The ${\it NB}816$, ${\it NB}921$, and ${\it NB}101$ images will be taken for the UD fields. The ${\it NB}387$, ${\it NB}816$, and ${\it NB}921$ observations will be conducted in 15 HSC-pointing D fields. 

Using the large HSC NB data complemented by optical and NIR spectroscopic observations, we launch a research project for Ly$\alpha$-emitting objects: {\it Systematic Identification of LAEs for Visible Exploration and Reionization Research Using Subaru HSC (SILVERRUSH)}. The large LAE samples provided by SILVERRUSH enable us to investigate e.g., LAE clustering (\cite{2017arXiv170407455O}), LEW LAEs and LABs (this work), spectroscopic properties of bright LAEs (\cite{shibuya2017b}), Ly$\alpha$ luminosity functions (\cite{2017arXiv170501222K}), and LAE overdensity (R. Higuchi et al. in preparation). The LAE survey strategy is given by \citet{2017arXiv170407455O}. This program is one of the twin programs. Another program is the study for dropouts, Great Optically Luminous Dropout Research Using Subaru HSC (GOLDRUSH), that is detailed in \citet{2017arXiv170406004O}, \citet{2017arXiv170406535H}, and \citet{2017arXiv170809421T}. 

\begin{figure}[tl!]
 \begin{center}
  \includegraphics[width=80mm]{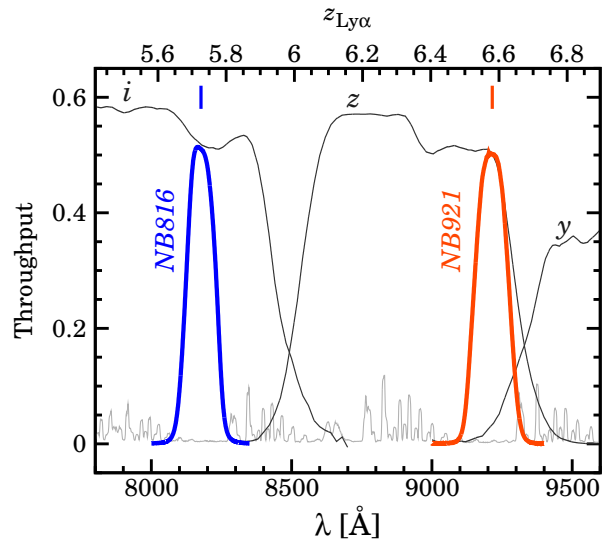}
 \end{center}
 \caption{Filter transmission curves of the NB and BB filters. The red and blue curves represent the {\it NB921} and {\it NB816} filters, respectively. The red and blue ticks show the NB central wavelengths with the same color coding as for the NB filter transmission curves. The black solid curves indicate the $i, z$, and $y$-band filters, from left to right. The gray line denotes the OH sky lines. The bandpass of these NB and BB filters corresponds to the area-weighted mean transmission curves\footnote{https://www.naoj.org/Observing/Instruments/HSC/sensitivity.html}. The transmission curves are derived by taking into account 1) the quantum efficiency of CCD, the transmittance of 2) the dewar window and 3) the HSC primary focus unit (POpt2), 4) the reflectivity of the primary mirror, and 5) the sky transparency (see \cite{2017arXiv170208449A}). The upper $x$-axis corresponds to the redshift of Ly$\alpha$. }\label{fig_hscnb}
\end{figure}

\begin{figure*}[tl!]
 \begin{center}
  \includegraphics[width=130mm]{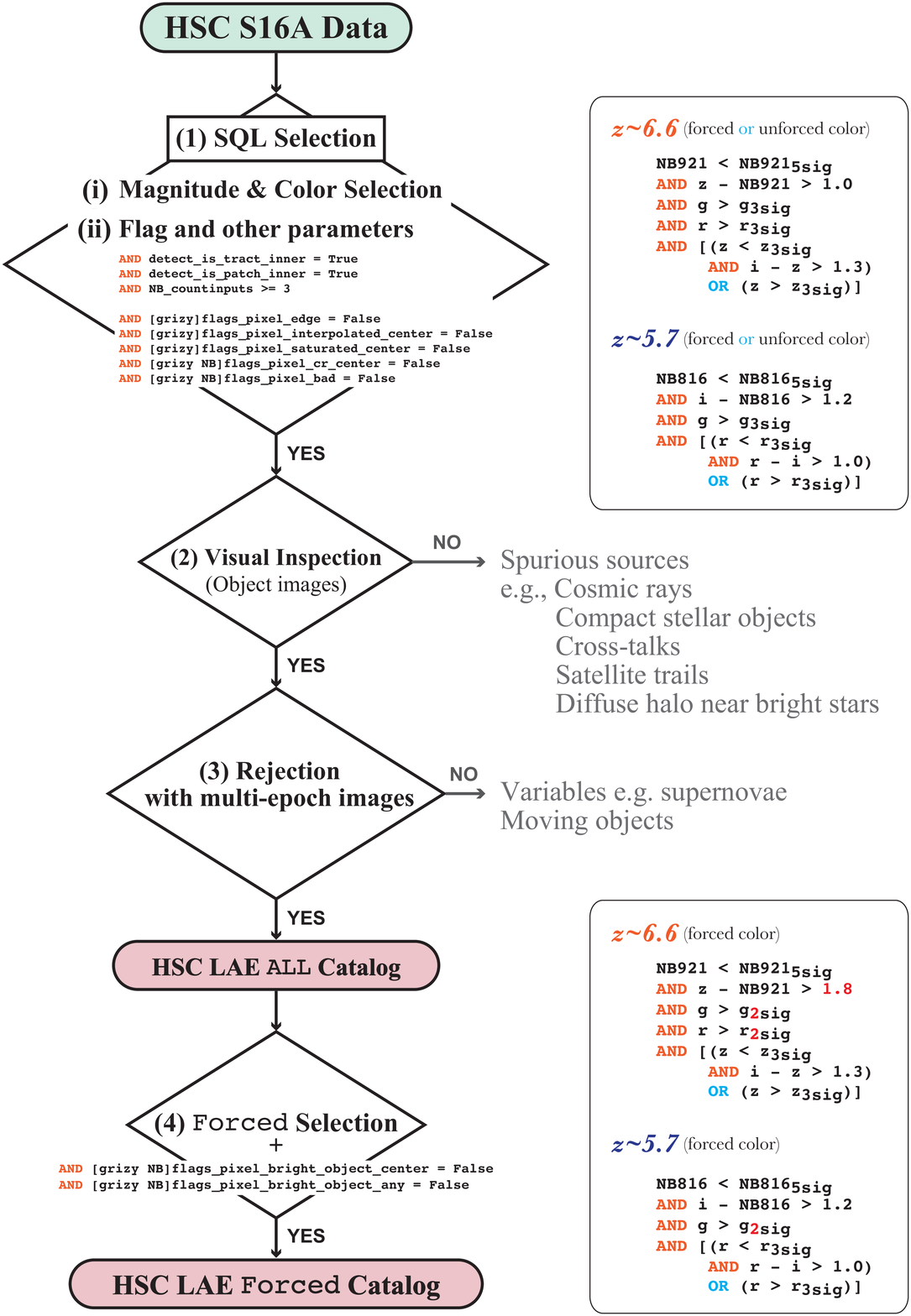}
 \end{center}
 \caption{Flow chart of the HSC LAE selection process. See Section \ref{sec_select} for more details. }\label{fig_hsclae_flowchart}
\end{figure*}

\begin{figure*}[t!]
 \begin{center}
  \includegraphics[width=150mm]{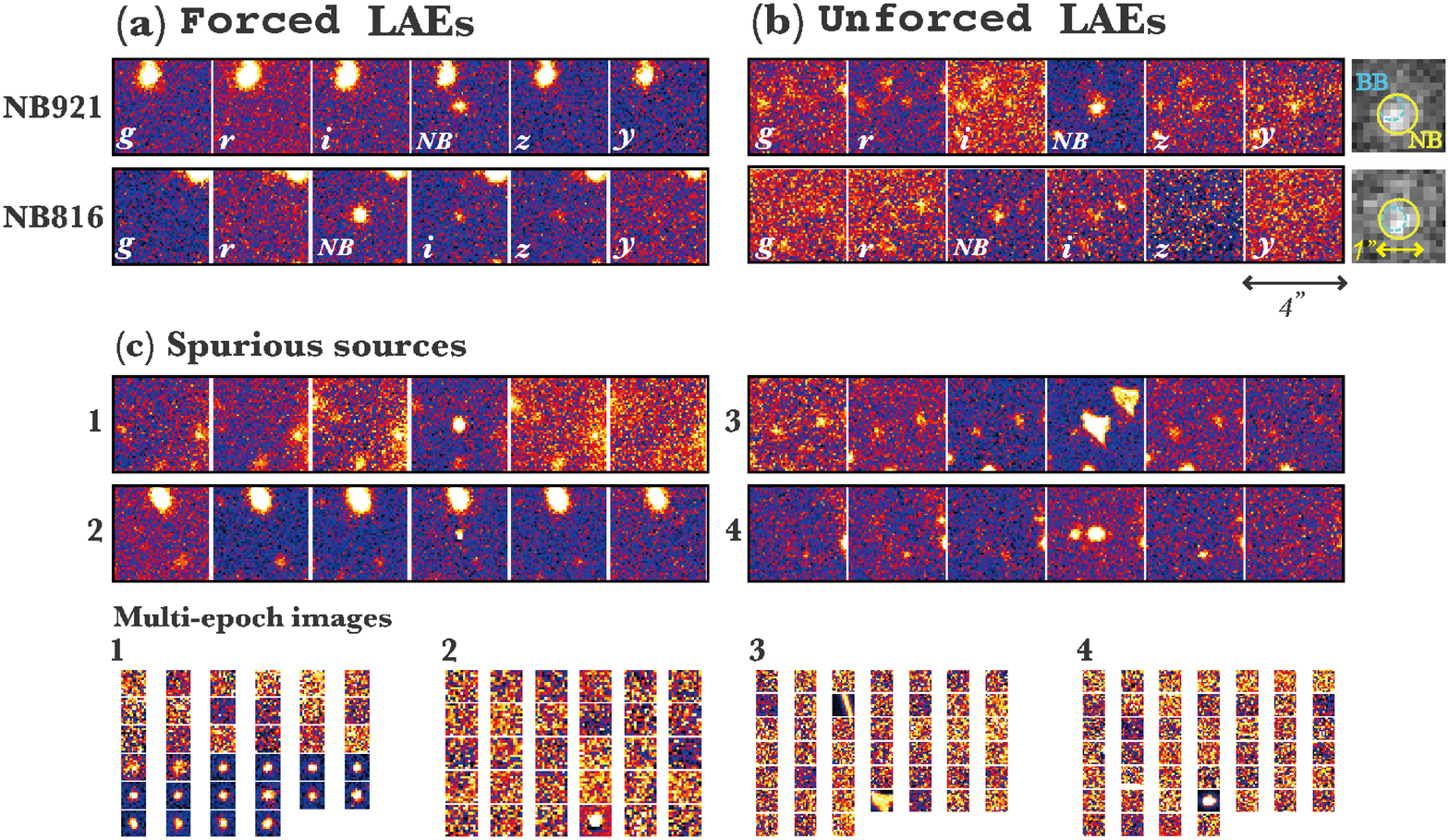}
 \end{center}
 \caption{Multi-band cutout images of our example LAEs and spurious sources. (a) LAEs at $z\simeq6.6$ (top) and $z\simeq5.7$ (bottom) in the {\tt forced} LAE catalog. (b) LAEs at $z\simeq6.6$ (top) and $z\simeq5.7$ (bottom) in the {\tt unforced} catalog. In the rightmost cutout images, the yellow solid and cyan dashed circles represent the central positions of the {\tt unforced} LAEs in the NB and BB images, respectively. The diameters of the yellow solid and dashed circles in the cutout images of the {\tt unforced} LAEs are $1^{\prime\prime}$ and $0.\!\!^{\prime\prime}5$, respectively. (c) Spurious sources with an NB magnitude-excess similar to that of LAE candidates (four panel sets at the top), 1: variable (e.g., supernova); 2: cosmic ray; 3: cross-talk artifact; 4: moving object (e.g., asteroids) and corresponding multi-epoch images (four panel sets at the bottom). The image size is $4^{\prime\prime}\times4^{\prime\prime}$ for the LAEs and spurious sources. }\label{fig_lae_examples}
\end{figure*}

This is the second paper in the SILVERRUSH project. In this paper, we present LAE selection processes and machine-readable catalogs of the LAE candidates at $z\simeq5.7-6.6$. Using the large LAE sample obtained with the first HSC NB data, we examine the redshift evolutions of Ly$\alpha$ EW distributions and LAB number density. This paper has the following structure. In Section \ref{sec_hscdata}, we describe the details of the SSP HSC data. Section \ref{sec_select} presents the LAE selection processes. In Section \ref{sec_check}, we check the reliability of our LAE selection. Section \ref{sec_results} presents Ly$\alpha$ EW distributions and LABs at $z\simeq6-7$. In Section \ref{sec_discuss}, we discuss the physical origins of LEW LAEs and LABs. We summarize our findings in Section \ref{sec_summary}.

Throughout this page, we adopt the concordance cosmology with $(\Omega_{\rm m}, \Omega_{\rm \Lambda}, h) = (0.3, 0.7, 0.7)$ (\cite{2016A&A...594A..13P}). All magnitudes are given in the AB system (\cite{1983ApJ...266..713O}). 

\section{HSC SSP Imaging Data}\label{sec_hscdata}

We use the HSC SSP S16A data products of $g, r, i, z$, and $y$ broadband (BB; \cite{kawanomoto2017}), ${\it NB}921$ and ${\it NB}816$ (\cite{2017arXiv170407455O}) images that are obtained in 2014-2016. It should be noted that this HSC SSP S16A data is significantly larger than the one of the first-data release in \citet{2017arXiv170208449A}. The ${\it NB}921$ (${\it NB}816$) filter has a central wavelength of $\lambda_c=9215$\AA\ ($8177$\AA) and an FWHM of $\Delta \lambda=135$\AA\ (113\AA), all of which are the area-weighted mean values. The ${\it NB}921$ and ${\it NB}816$ filters trace the redshifted Ly$\alpha$ emission lines at $z=6.580\pm0.056$ and $z=5.726\pm0.046$, respectively. The NB filter transmission curves are shown in Figure \ref{fig_hscnb}. The central wavelength, FWHM, and the bandpass shape for these NB filters are almost uniform over the HSC FoV. The deviation of the $\lambda_c$ and FWHM values are typically within $\simeq0.3$\% and $\simeq10$\%, respectively. Thus, we use the area-weighted mean transmission curves in this study. The detailed specifications of these NB filters are given in \citet{2017arXiv170407455O}.

Table \ref{tab_nbdata} summarizes the survey areas, exposure time, and depth of the HSC SSP S16A NB data. The current HSC SSP S16A NB data covers UD-COSMOS, UD-SXDS, D-COSMOS, D-DEEP2-3, D-ELAIS-N1 for $z\simeq6.6$, and UD-COSMOS, UD-SXDS, D-DEEP2-3, D-ELAIS-N1 for $z\simeq5.7$. The effective survey areas of the ${\it NB}921$ and ${\it NB}816$ images are $21.2$ and $13.8$ arcmin$^2$, corresponding to the survey volumes of $\simeq1.9\times 10^7$ and $\simeq1.2\times 10^7$ Mpc$^3$, respectively. The area of these HSC NB fields are covered by the observations of all the BB filters. The typical limiting magnitudes of BB filters are $g\simeq26.9$, $r\simeq26.5$, $r\simeq26.3$, $z\simeq25.7$, and $y\simeq25.0$ ($g\simeq26.6$, $r\simeq26.1$, $r\simeq25.9$, $z\simeq25.2$, and $y\simeq24.4$) in a $1.\!\!\arcsec5$ aperture at $5\sigma$ for the UD (D) fields. The FWHM size of point spread function in the HSC images is typically $\simeq0.\!\!\arcsec8$ (\cite{2017arXiv170208449A}). 

The HSC images were reduced with the HSC pipeline, {\tt hscPipe} 4.0.2 (\cite{2017arXiv170506766B}) which is a code from the Large Synoptic Survey Telescope (LSST) software pipeline (\cite{2008arXiv0805.2366I, doi:10.1117/12.857297, 2015arXiv151207914J}). The HSC pipeline performs CCD-by-CCD reduction, calibration for astrometry, and photometric zero point determination. The pipeline then conducts mosaic-stacking that combines reduced CCD images into a large coadd image, and create source catalogs by detecting and measuring sources on the coadd images. The photometric calibration is carried out with the PanSTARRS1 processing version 2 imaging survey data (\cite{2013ApJS..205...20M,2012ApJ...756..158S,2012ApJ...750...99T}). The details of the HSC SSP survey, data reduction, and source detection and photometric catalog construction are provided in \citet{2017arXiv170208449A}, \citet{2017arXiv170405858A}, and \citet{2017arXiv170506766B}. 

In the HSC images, source detection and photometry were carried out in two methods: {\tt unforced} and {\tt forced}. The {\tt unforced} photometry is a method to perform measurements of coordinates, shapes, and fluxes individually in each band image for an object. The {\tt forced} photometry is a method to carry out photometry by fixing centroid and shape determined in a reference band and applying them to all the other bands. The algorithm of the {\tt forced} detection and photometry is similar to the {\tt double-image} mode of {\tt SExtractor} (\cite{1996A&AS..117..393B}) that are used in most of the previous studies for high-$z$ galaxies. According to which depends on magnitudes, $S/N$, positions, and profiles for detected sources, one of the BB and NB filter is regarded as a reference band. For merging the catalogs of each band, the object matching radius is not a specific value which depends on an area of regions with a $>5\sigma$ sky noise level. We refer the detailed algorithm to choose the reference filter and filter priority to \citet{2017arXiv170506766B}. 

In the {\tt hscPipe} detection and photometry, an NB filter is basically chosen as a reference band for the NB-bright and BB-faint sources such as LAEs. However, a BB filter is used as a reference band in the case that sources are bright in the BB image. The current version of {\tt hscPipe} has not implemented the NB-reference {\tt forced} photometry for BB-bright sources. In this specification, there is a possibility that we miss BB-bright sources with a spatial offset between centroids of BB and NB by using only the {\tt forced} photometry. Thus, we combine the {\tt unforced} or {\tt forced} photometry for $BB-NB$ colors to identify such BB-bright objects with a spatial offset between centroids of BB and NB (e.g., \cite{2014arXiv1401.1209S}). See Section \ref{sec_select} for details of the LAE selection criteria. 

We use {\tt cmodel} magnitudes for estimating total magnitudes of sources. The cmodel magnitude is a weighted combination of exponential and de Vaucouleurs fits to light profiles of each object. The detailed algorithm of the cmodel photometry are presented in \citet{2017arXiv170506766B}. To measure the $S/N$ values for source detections, we use $1.\!\!^{\prime\prime}5$-diameter aperture magnitudes.

\begin{longtable}{*{8}{c}}
\caption{Properties of the HSC SSP S16A NB Data}\label{tab_nbdata}
\hline
Field & R.A. & Dec. & Area & $T_{\rm exp}$ & $m_{\rm lim} (5\sigma, 1.5^{\prime\prime}\phi)$ & $N_{\rm LAE,ALL}$ & $N_{\rm LAE,F}$  \\
& (J2000) & (J2000) & (deg$^2$) & (hour) & (mag) & & \\
(1) & (2) & (3) & (4) & (5) & (6) & (7) & (8) \\
\hline
\endfirsthead
\endhead
\hline
\endfoot
\hline
\multicolumn{8}{l}{(1) Field.} \\
\multicolumn{8}{l}{(2) Right ascension. } \\
\multicolumn{8}{l}{(3) Declination.} \\
\multicolumn{8}{l}{(4) Survey area with the HSC SQL parameters in Table \ref{tab_flag}. } \\
\multicolumn{8}{l}{(5) Total exposure time of the NB imaging observation.} \\
\multicolumn{8}{l}{(6) Limiting magnitude of the NB image defined by a $5\sigma$ sky noise in a $1.\!\!^{\prime\prime}5$ diameter circular aperture. } \\
\multicolumn{8}{l}{(7) Number of the LAE candidates in the {\tt ALL} ({\tt unforced}+{\tt forced}) catalog. } \\
\multicolumn{8}{l}{(8) Number of the LAE candidates in the {\tt forced} catalog. } \\
\multicolumn{8}{l}{$^a$ The value of $N_{\rm LAE,ALL}$ ($N_{\rm LAE,F}$) includes 30 (7) LAEs selected in UD-COSMOS.} \\
\endlastfoot
\multicolumn{8}{c}{${\it NB}921$ ($z\simeq6.6$)} \\ \hline 
UD-COSMOS & 10:00:28 & $+$02:12:21 & 2.05 & 11.25 & 25.6 & 338 & 116 \\
UD-SXDS & 02:18:00 & $-$05:00:00 & 2.02 & 7.25 & 25.5 & 58 & 23 \\
D-COSMOS & 10:00:60 & $+$02:13:53 & 5.31 & 2.75 & 25.3 & 244$^a$ & 47$^a$ \\
D-DEEP2-3 & 23:30:22 & $-$00:44:38 & 5.76 & 1.00 & 24.9 & 164 & 35 \\
D-ELAIS-N1 & 16:10:00 & $+$54:17:51 & 6.08 & 1.75 & 25.3 & 349 & 48 \\ 
{\bf Total} & --- & --- & 21.2 & 24.00 & --- & 1153 & 269 \\ \hline 
\multicolumn{8}{c}{${\it NB}816$  ($z\simeq5.7$)} \\ \hline 
UD-COSMOS & 10:00:28 & $+$02:12:21 & 1.97 & 5.50 & 25.7 & 201 & 176 \\
UD-SXDS & 02:18:00 & $-$05:00:00 & 1.93 & 3.75 & 25.5 & 224 & 188 \\
D-DEEP2-3 & 23:30:22 & $-$00:44:38 & 4.37 & 1.00 & 25.2 & 423 & 282 \\
D-ELAIS-N1 & 16:10:00 & $+$54:17:51 & 5.56 & 1.00 & 25.3 & 229 & 130 \\ 
{\bf Total} & --- & --- & 13.8 & 11.25 & --- & 1077 & 776 \\ \hline
\end{longtable}

\begin{longtable}{llcl}
\caption{HSC SQL Parameters and Flags for Our LAE Selection}\label{tab_flag}
\hline
Parameter or Flag & Value & Band & Comment  \\
\hline
\endfirsthead
\endhead
\hline
\endfoot
\hline
\multicolumn{4}{l}{} \\
\endlastfoot
{\tt detect\_is\_tract\_inner} & {\tt True} & --- & Object is in an inner region of a tract and \\
 & & & not in the overlapping region with adjacent tracts \\
 {\tt detect\_is\_patch\_inner} & {\tt True} & --- & Object is in an inner region of a patch and \\
  & & & not in the overlapping region with adjacent patches \\
 {\tt countinputs} & $>=3$ & {\it NB} & Number of visits at a source position for a given filter. \\ 
{\tt flags\_pixel\_edge} & {\tt False} & $grizy$, {\it NB} & Locate within images \\
{\tt flags\_pixel\_interpolated\_center}  & {\tt False} & $grizy$, {\it NB} & None of the central $3\times3$ pixels of an object is interpolated \\
{\tt flags\_pixel\_saturated\_center}  & {\tt False} & $grizy$, {\it NB} & None of the central $3\times3$ pixels of an object is saturated \\
{\tt flags\_pixel\_cr\_center}  & {\tt False} & $grizy$, {\it NB} & None of the central $3\times3$ pixels of an object is masked as cosmic ray \\
{\tt flags\_pixel\_bad}  & {\tt False} & $grizy$, {\it NB} & None of the pixels in the footprint of an object is labelled as bad \\ 
\hline
\end{longtable}

\begin{figure*}[t!]
 \begin{center}
  \includegraphics[width=150mm]{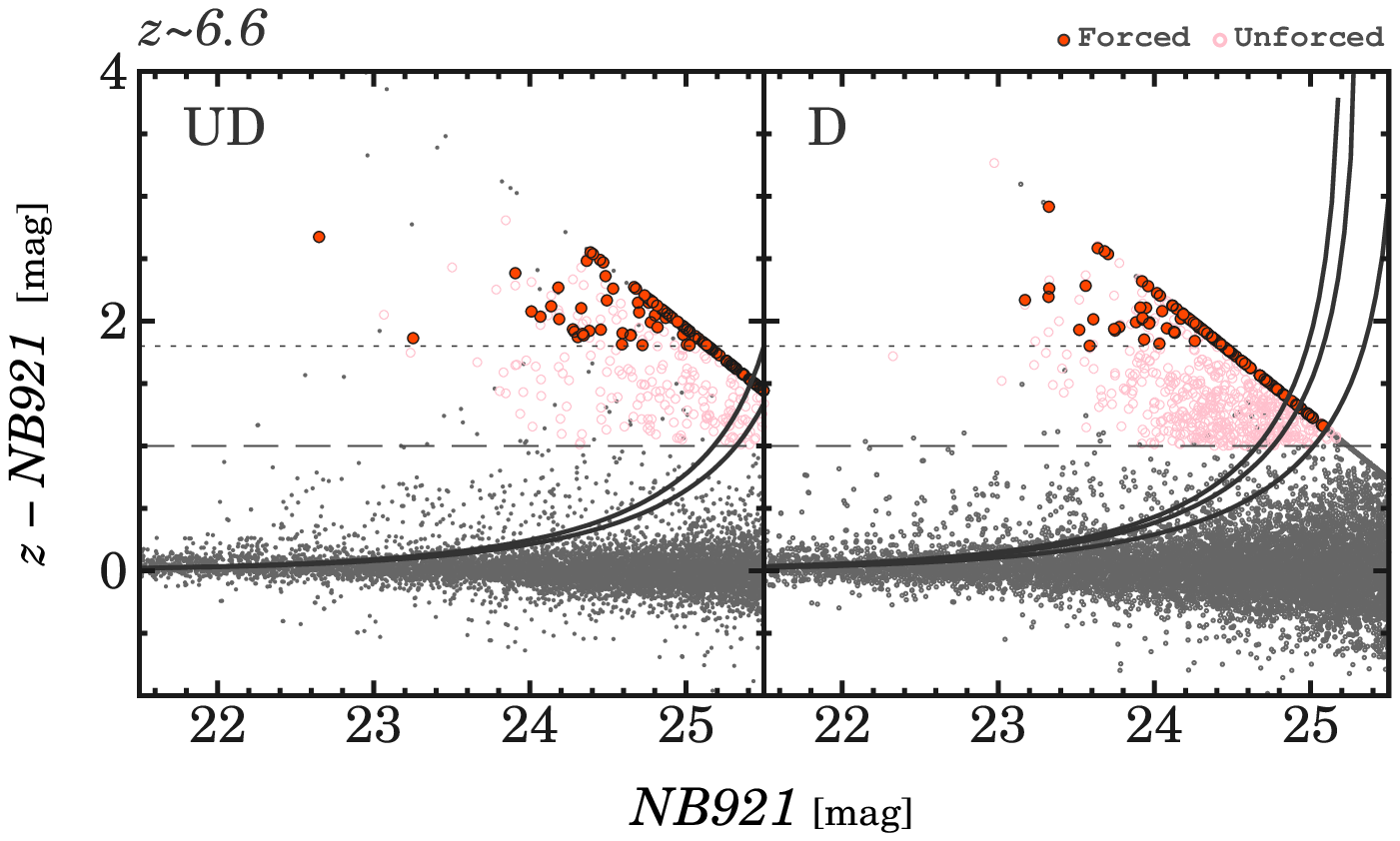}
  \includegraphics[width=150mm]{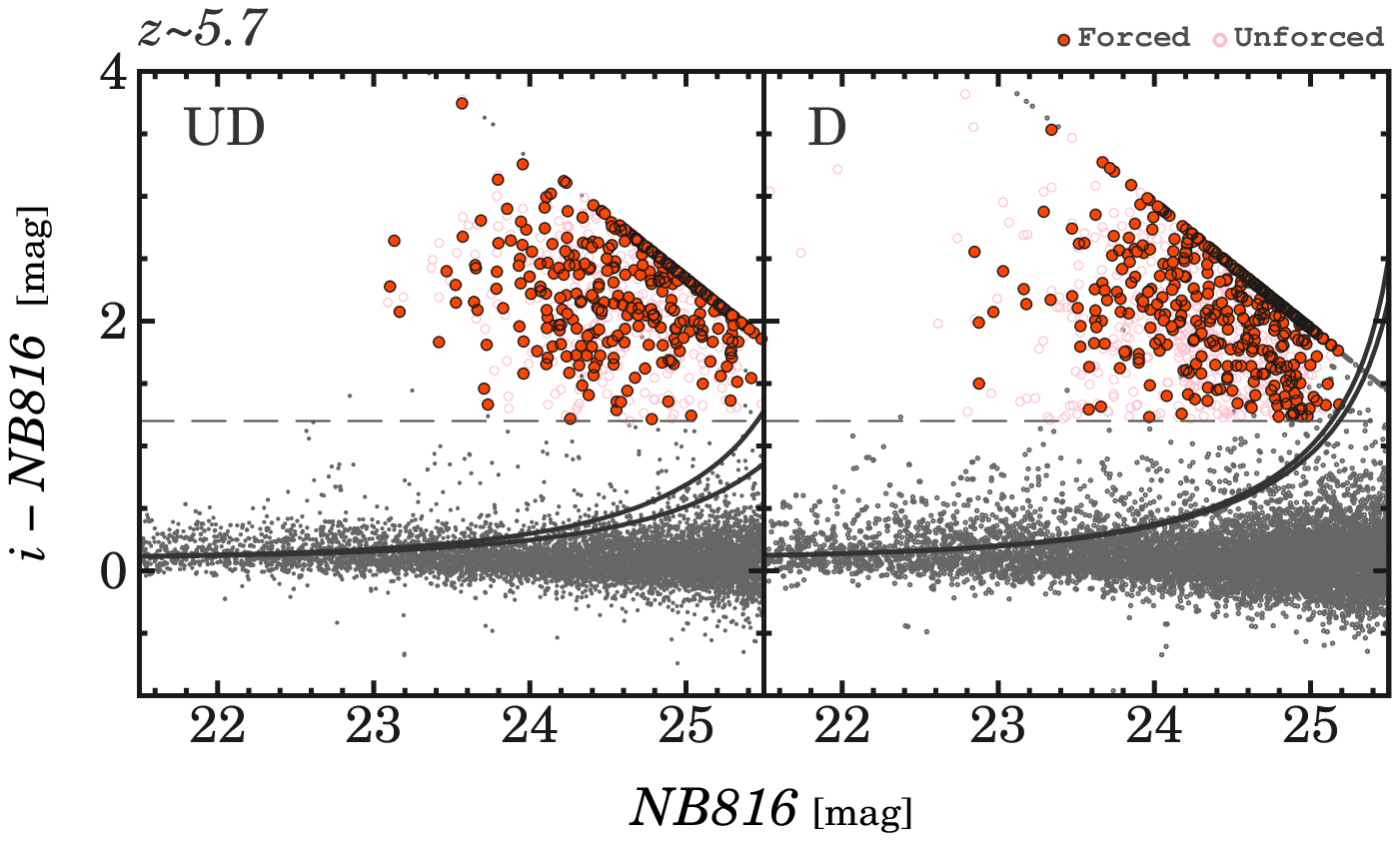}
 \end{center}
 \caption{({\bf Top}) Color of $z-{\it NB}921$ as a function of ${\it NB}921$ magnitude for the LAEs at $z\simeq6.6$ in the UD (left) and D (right) fields. The filled red and open magenta circles denote the {\tt forced} and {\tt unforced} LAEs, respectively. For the LAEs undetected in the $z$-band images, the $z$-band magnitudes are replaced with the $2\sigma$ limiting magnitudes. The x-axis denotes the {\tt forced} ({\tt unforced}) $z-{\it NB}921$ colors for the {\tt forced} ({\tt unforced}) LAEs. The horizontal dashed and dotted line shows the color criteria of $z-{\it NB}921>1.0$ and $z-{\it NB}921>1.8$, respectively. The gray dots present objects detected in the ${\it NB}921$ images. The solid curves show the $3\sigma$ error tracks of $z-{\it NB}921$ color for each field. The $3\sigma$ error tracks are derived by Equation \ref{eq_limit}. ({\bf Bottom}) Color of $i-{\it NB}816$ as a function of ${\it NB}816$ magnitude for the LAEs at $z\simeq5.7$. The definitions of symbols, curves, and lines are the same as those of the top panels. }\label{fig_nb_color_tile}
\end{figure*}

\begin{figure*}[t!]
 \begin{center}
  \includegraphics[width=170mm]{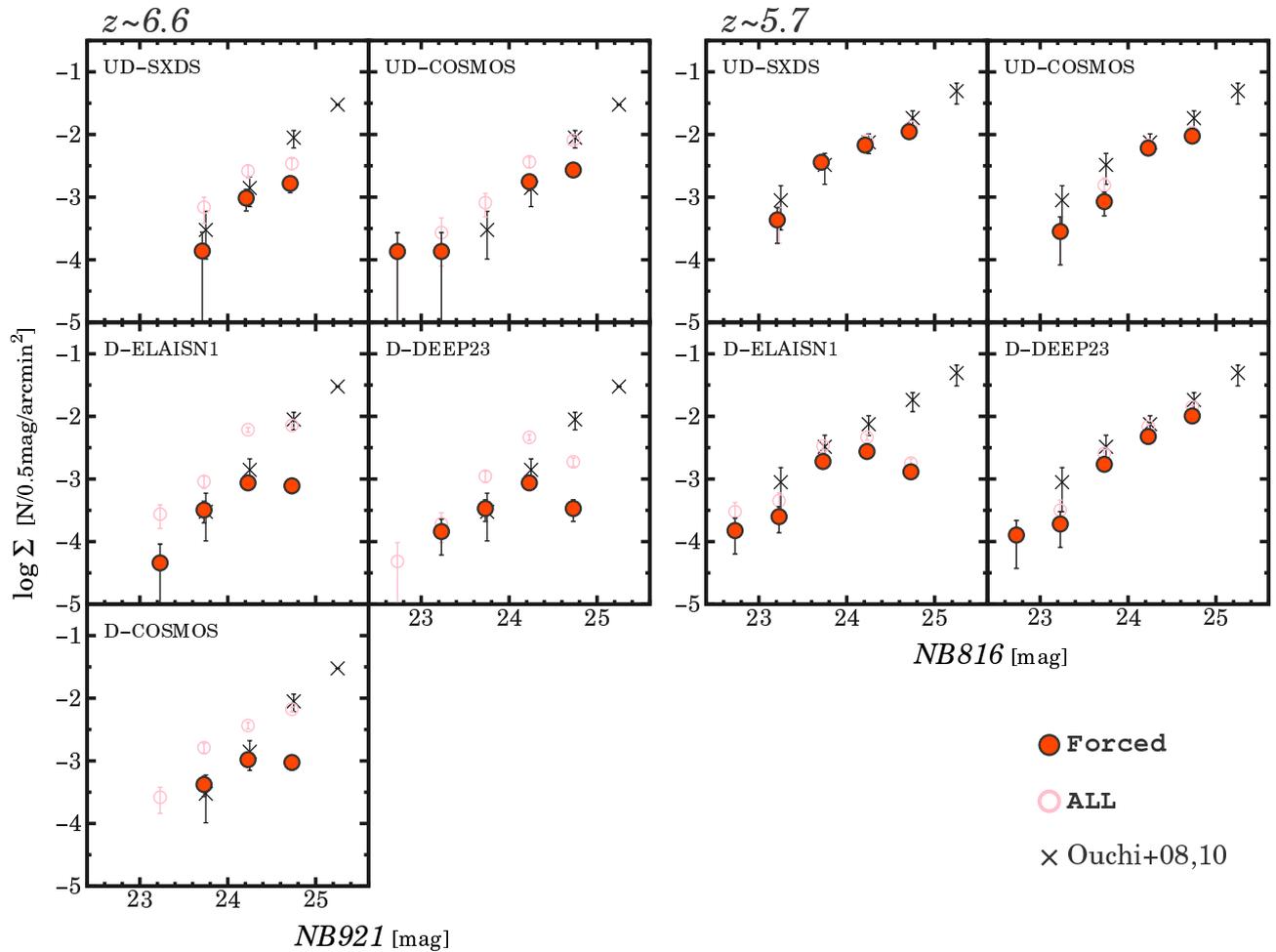}
 \end{center}
 \caption{Surface number density (SND) of the HSC LAEs at $z\simeq6.6$ (five panels at the left) and $\simeq5.7$ (four panels at the right) in each UD and D field. The filled red and open magenta circles indicate the LAEs in the {\tt forced} and {\tt ALL} catalog, respectively. The error bars are given by Poisson statistics from the number of LAEs. The gray crosses represent the LAEs in \citet{2010ApJ...723..869O} for $z\simeq6.6$ and \citet{2008ApJS..176..301O} for $z\simeq5.7$. The data points of the gray crosses are identical in all the fields for each redshift. The SND slight declines in the HSC LAEs at ${\it NB}\gtrsim24.5$ mag would be originated from the incompleteness of the LAE detection and selection. The completeness-corrected SNDs are presented in \citet{2017arXiv170501222K}. The data points of the HSC LAEs are slightly shifted along x-axis for clarity. }\label{fig_nb_sd_mf_tile}
\end{figure*}

\begin{figure*}[t!]
 \begin{center}
  \includegraphics[width=150mm]{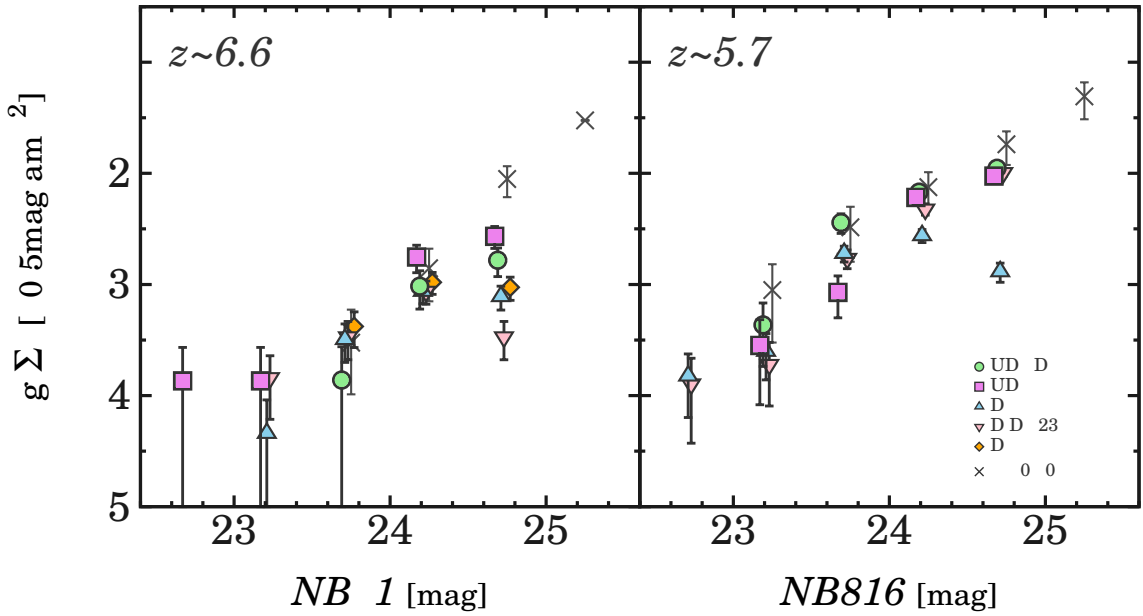}
 \end{center} \caption{Surface number density (SND) as a function of NB magnitude for the LAEs at $z\sim6.6$ (left) and $\sim5.7$ (right) in the HSC LAE {\tt forced} catalog. The colored symbols denote the LAEs in each UD and D field (green circles: UD-SXDS; magenta squares: UD-COSMOS; cyan triangles: D-ELAIS-N1; light-red inverse-triangles: D-DEEP2-3; orange diamonds: D-COSMOS). The error bars are given by Poisson statistics from the number of LAEs. The gray crosses represent the LAEs in \citet{2010ApJ...723..869O} for $z\simeq6.6$ and \citet{2008ApJS..176..301O} for $z\simeq5.7$. The SND slight declines in the HSC LAEs at ${\it NB}\gtrsim24.5$ mag would be originated from the incompleteness of the LAE selection. The completeness-corrected SNDs are presented in \citet{2017arXiv170501222K}. The data points of the HSC LAEs are slightly shifted along x-axis for clarity. }\label{fig_nb_sd_tile}
\end{figure*}

\begin{longtable}{*{7}{c}}
\caption{Photometric properties of example LAE candidates}\label{tab_catalog}
\hline
Object ID & ${\it NB}$ & $g$ & $r$ & $i$ & $z$ & $y$ \\
 & (mag) & (mag) & (mag) & (mag) & (mag) & (mag) \\
(1) & (2) & (3) & (4) & (5) & (6) & (7) \\
\hline
\endfirsthead
\endhead
\hline
\endfoot
\hline
\multicolumn{7}{l}{(1) Object ID.} \\
\multicolumn{7}{l}{(2)-(7) Total magnitude of {\it NB}-, {\it g}-, {\it r}-, {\it i}-, {\it z}, and {\it y}-bands. } \\
\multicolumn{7}{l}{The 2$\sigma$ limits of the total magnitudes for the undetected bands. } \\
\multicolumn{7}{l}{(The complete machine-readable catalogs will be available on our project webpage at} \\
\multicolumn{7}{l}{ http://cos.icrr.u-tokyo.ac.jp/rush.html.) } \\
\endlastfoot
\multicolumn{7}{c}{UD-SXDS (${\it NB}921$)} \\ \hline 
HSC J021601$-$041442 & $23.85\pm0.10$ & $26.89\pm0.45$ & $27.03\pm0.62$ & $26.65\pm0.63$ & $25.28\pm0.31$ & $25.29\pm0.53$ \\ 
HSC J021754$-$051454 & $24.01\pm0.12$ & $>27.6$ & $>27.3$ & $>26.9$ & $26.09\pm0.57$ & $25.21\pm0.50$ \\ 
HSC J021702$-$050604 & $24.64\pm0.21$ & $>27.6$ & $>27.3$ & $>26.9$ & $>26.5$ & $>25.8$ \\ 
HSC J021638$-$043228 & $24.74\pm0.23$ & $>27.6$ & $>27.3$ & $>26.9$ & $26.17\pm0.60$ & $>25.8$ \\ 
HSC J021609$-$050236 & $24.90\pm0.26$ & $27.53\pm0.72$ & $27.29\pm0.75$ & $>26.9$ & $26.32\pm0.67$ & $>25.8$ \\ 
\hline
\multicolumn{7}{c}{UD-COSMOS (${\it NB}816$)} \\ \hline 
HSC J100243$+$024551 & $23.69\pm0.08$ & $>27.6$ & $>27.3$ & $26.49\pm0.53$ & $>26.6$ & $>25.8$ \\ 
HSC J100239$+$022806 & $24.14\pm0.13$ & $>27.6$ & $>27.3$ & $26.76\pm0.64$ & $26.12\pm0.54$ & $>25.8$ \\ 
HSC J100243$+$015931 & $24.63\pm0.19$ & $>27.6$ & $>27.3$ & $>27.0$ & $>26.6$ & $>25.8$ \\ 
HSC J095936$+$014108 & $25.02\pm0.26$ & $>27.6$ & $>27.3$ & $>27.0$ & $>26.6$ & $>25.8$ \\ 
HSC J100245$+$021536  & $25.15\pm0.29$ & $>27.6$ & $>27.3$ & $>27.0$ & $>26.6$ & $>25.8$ \\ 
\end{longtable}

\section{LAE Selection}\label{sec_select}

Using the HSC data, we perform a selection for LAEs at $z\simeq6.6$ and $\simeq5.7$. Basically, we select objects showing a significant flux excess in the NB images and a spectral break at the wavelength of redshifted Ly$\alpha$ emission. In this study, we create two LAE catalogs: {\it HSC LAE {\tt ALL (forced+unforced)} catalog} and {\it HSC LAE {\tt forced} catalog}. The HSC LAE {\tt ALL} catalog is constructed in a combination of the {\tt forced} and {\tt unforced} photometry. We use this HSC LAE {\tt ALL} catalog for identifying objects with a spatial offset between centroids of BB and NB (see Section \ref{sec_hscdata}). On the other hand, the HSC LAE {\tt forced} catalog consists of LAEs meeting only the selection criteria of the {\tt forced} photometry. We use this HSC LAE {\tt forced} catalog for statistical studies for LAEs (e.g., Ly$\alpha$ LFs). The HSC LAE {\tt forced} catalog is a subsample of the {\tt ALL} one. Figure \ref{fig_hsclae_flowchart} shows the flow chart of the LAE selection process. We carry out the following processes: (1) SQL selection, (2) visual inspections for the object images, (3) rejections of variable and moving objects with the multi-epoch images, and (4) {\tt forced} selection. The details are described as below.

\begin{enumerate}
\renewcommand{\labelenumi}{(\arabic{enumi})}
\item {\bf SQL selection: } We retrieve detection and photometric catalogs from postgreSQL database tables.  Using SQL scripts, we select objects meeting the following criteria of (i) magnitude and color selections and (ii) {\tt hscPipe} parameters and flags. 
\begin{enumerate}
\renewcommand{\labelenumii}{(\roman{enumii})}
\item {\bf Magnitude and color selection: } To identify objects with an ${\it NB}$ magnitude excess in the HSC catalog, we apply the magnitude and color selection criteria that are similar to e.g., \cite{2008ApJS..176..301O,2010ApJ...723..869O}: 

\begin{equation}
\begin{array}{l}
\textit{NB921}^\mathrm{ap}_\mathrm{frc} < \textit{NB921}_{5\sigma}  \\
\&\& \ \left(\textit{g}_\mathrm{frc} > \textit{g}_{3\sigma}  \ || \ \textit{g}^\mathrm{ap}_\mathrm{frc} > \textit{g}_{3\sigma} \right)  \\
\&\& \ \left(\textit{r}_\mathrm{frc} > \textit{r}_{3\sigma}  \ || \ \textit{r}^\mathrm{ap}_\mathrm{frc} > \textit{r}_{3\sigma} \right)  \\
\&\& \ \left( \textit{z}_\mathrm{frc} - \textit{NB921}_\mathrm{frc} > 1.0 \ || \ \textit{z}_\mathrm{unf} - \textit{NB921}_\mathrm{unf} > 1.0 \right) \\
\&\& \ \left\{ \ \left[  (\textit{z}_\mathrm{frc} < \textit{z}_{3\sigma}  \ || \ \textit{z}^\mathrm{ap}_\mathrm{frc} < \textit{z}_{3\sigma}) \right. \right. \\
\left. \left. \ \ \ \ \ \ \ \ \&\& \ (\textit{i}_\mathrm{frc} - \textit{z}_\mathrm{frc} > 1.3  \ || \  \textit{i}_\mathrm{unf} - \textit{z}_\mathrm{unf} > 1.3) \right] \right.    \\
\left. \ \ \ \ \ \ \ \ || \ (\textit{z}_\mathrm{frc} > \textit{z}_{3\sigma}  \ || \ \textit{z}^\mathrm{ap}_\mathrm{frc} > \textit{z}_{3\sigma}) \ \right\}, \\
\end{array}
\label{eq_6p6select}
\end{equation}

\mbox{} \\
for $z\simeq6.6$, and, \\ 
\mbox{}

\begin{equation}
\begin{array}{l}
\textit{NB816}^\mathrm{ap}_\mathrm{frc} < \textit{NB816}_{5\sigma}  \\
\&\& \ \left(\textit{g}_\mathrm{frc} > \textit{g}_{3\sigma}  \ || \ \textit{g}^\mathrm{ap}_\mathrm{frc} > \textit{g}_{3\sigma} \right)  \\
\&\& \ \left( \textit{i}_\mathrm{frc} - \textit{NB816}_\mathrm{frc} > 1.2 \ || \ \textit{i}_\mathrm{unf} - \textit{NB816}_\mathrm{unf} > 1.2 \right)   \\
\&\& \ \left\{ \ \left[  (\textit{r}_\mathrm{frc} < \textit{r}_{3\sigma}  \ || \ \textit{r}^\mathrm{ap}_\mathrm{frc} < \textit{r}_{3\sigma}) \right. \right. \\
\left. \left. \ \ \ \ \ \ \ \ \&\& \ (\textit{r}_\mathrm{frc} - \textit{i}_\mathrm{frc} > 1.0  \ || \  \textit{r}_\mathrm{unf} - \textit{i}_\mathrm{unf} > 1.0) \right] \right.    \\
\left. \ \ \ \ \ \ \ \ || \ (\textit{r}_\mathrm{frc} > \textit{r}_{3\sigma}  \ || \ \textit{r}^\mathrm{ap}_\mathrm{frc} > \textit{r}_{3\sigma}) \ \right\},
\end{array}
\label{eq_5p7select}
\end{equation}

\mbox{}

\noindent for $z\simeq5.7$, where the indices of {\tt frc} and {\tt unf} represent the forced and unforced photometry, respectively. The subscript of $5\sigma$ ($3\sigma$) indicates the $5\sigma$ ($3\sigma$) limiting magnitude for a given filter. The values with and without a superscript of {\tt ap} indicate the aperture and total magnitudes, respectively. These magnitudes are derived with the {\tt hscPipe} software (see Section \ref{sec_hscdata}; \cite{2017arXiv170506766B}). The limits of the $i-NB816$ and $z-NB921$ colors are the same as those of \citet{2008ApJS..176..301O} and \citet{2010ApJ...723..869O}, respectively. To exploit the survey capability of HSC identifying rare objects, we use the $3\sigma$ $g$ and $r$ limiting magnitude (instead of the value of $2\sigma$ used in \cite{2008ApJS..176..301O}) for the criteria of Lyman break off-band non-detection. In the process (4), we replace $3\sigma$ with $2\sigma$ for the $g$ and $r$ magnitude criteria for the consistency with the previous studies. 

Note that we do not apply the {\tt flags\_pixel\_bright\_object\_[center/any]} masking to the LAE {\tt ALL} catalog in order to maximize LAE targets for future follow-up observations (\cite{2017arXiv170208449A}). These flags for the object masking are used in the process (4).

\item {\bf Parameters and flags: } Similar to \citet{2017arXiv170406004O}, we set several {\tt hscPipe} parameters and flags in the HSC catalog to exclude e.g., blended sources, and objects affected by saturated pixels, and nearby bright source halos. We also mask regions where exposure times are relatively short by using the {\tt countinputs} parameter, $N_{\rm c}$, which denotes the number of exposures at a source position for a given filter. Table \ref{tab_flag} summarizes the values and brief explanations of the {\tt hscPipe} parameters and flags used for our LAE selection. The full details of these parameters and flags are presented in \citet{2017arXiv170208449A}. To search for LAEs in large areas of the HSC fields, we do not apply the {\tt countinputs} parameter to the BB images.

\end{enumerate}

The number of objects selected in this process is $n_{\rm SQL} \simeq121,000$. 

\item {\bf Visual inspections for object images: } To exclude cosmic rays, cross-talks, compact stellar objects, and artificial diffuse objects, we perform visual inspections for the BB and NB images of all the objects selected in the process (1). Most spurious sources are diffuse components near bright stars and extended nearby galaxies. The {\tt hscPipe} software conducts the cmodel fit to broad light profiles of such diffuse sources in the NB images, which enhances the $BB-NB$ colors. For this reason, the samples constructed in the current {\tt SQL} selection are contaminated by many diffuse components. Due to the clear difference of the appearance between LAE candidates and diffuse components, such spurious sources can be easily excluded through the visual inspections. The number of objects selected in this process is $n_{\rm vis} \simeq 10,900$.

The visual inspection processes are mainly conducted by one of the authors. For the reliability check, four authors in this paper have individually carried out such visual inspections for $\simeq5,300$ objects in the UD-COSMOS ${\it NB}816$ fields, and compare the results of the LAE selection. The difference in the number of selected LAEs is within $\pm5$ objects. Thus, we do not find a large difference in our visual inspection results.

\item {\bf Rejection of variable and moving objects with multi-epoch images: } We exclude variable and moving objects such as supernovae, AGNs, satellite trails, and asteroids using multi-epoch NB images. The NB images were typically taken a few months - years after the BB imaging observations. For this reason, there is a possibility that sources with an NB flux excess are variable or moving objects which happened to enhance the luminosities during the NB imaging observations. 

The NB images are created by coadding $\simeq10-20$ and $\simeq3-5$ frames of $15$ minute exposures for the current HSC UD and D data, respectively. Using the multi-epoch images, we automatically remove the variable and moving objects as follows. First, we measure the flux for individual epoch images, $f_{\rm 1 epoch}$, for each object. Next, we obtain an average, $f_{\rm ave}$, and a standard deviation, $\sigma_{\rm epoch}$, from a set of the $f_{\rm 1 epoch}$ values after a $2\sigma$ flux clipping. Finally, we discard an object having at least a multi-epoch image with a significantly large $f_{\rm 1 epoch}$ value of $f_{\rm 1 epoch} \geq f_{\rm ave}+ A_{\rm epoch} \times \sigma_{\rm epoch}$. Here we tune the $A_{\rm epoch}$ factor based on the depth of the NB fields. The $A_{\rm epoch}$ value is typically $\simeq2.0-2.5$. Figure \ref{fig_lae_examples} shows examples of the spurious sources. 

We also perform visual inspections for multi-epoch images to remove contaminants which are not excluded in the automatic rejection above. We refer the remaining objects after this process as the LAE {\tt ALL} catalog. 

\item {\bf {\tt Forced} selection: } 

In the selection criteria of Equations (\ref{eq_6p6select}) and (\ref{eq_5p7select}), the HSC LAE {\tt ALL} catalog is obtained in the combination of the {\tt forced} and {\tt unforced} colors. In this process, we select LAEs only with the {\tt forced} color excess to create the {\tt forced} LAE subsamples from the HSC LAE {\tt ALL} catalog. In addition, the $3\sigma$ limit is replaced with $2\sigma$ for the criteria of $g$ and $r$ band non-detections.  

Here we also adopt a new stringent color criterion of $z-{\it NB}921>1.8$ for $z\simeq6.6$ LAEs. Due to the difference of the $z$ band transmission curves between SCam and HSC, the criterion of $z-{\it NB}921>1.0$ in Equation (\ref{eq_6p6select}) do not allow us to select LAEs whose $EW_{\rm 0, Ly\alpha}$ is similar to those of previous SCam studies. The $BB-NB$ color criteria in in the {\tt forced} selection correspond to the rest-frame Ly$\alpha$ EW of $EW_{\rm 0, Ly\alpha} > 14$\,\AA\ and $> 10$\,\AA\ for $z\simeq6.6$ and $z\simeq5.7$ LAEs, respectively. These $EW_{\rm 0,Ly\alpha}$ limits are comparable to those of the previous SCam studies (e.g., \cite{2010ApJ...723..869O}). The relation between $EW_{\rm 0,Ly\alpha}$ and $BB-NB$ colors is described in \citet{2017arXiv170501222K} in details. Moreover, we remove the objects in masked regions defined by the {\tt flags\_pixel\_bright\_object\_[center/any]} parameters (\cite{2017arXiv170208449A}). 

We refer the set of the remaining objects after this process as {\it the {\tt forced} LAE catalog}. This {\tt forced} LAE catalog is used for studies on LAE statistics such as measurements of Ly$\alpha$ EW scale lengths.

The LAE candidates selected in this {\tt forced} selection are referred to as the {\tt forced} LAEs. On the other hand, we refer to the remaining LAE candidates in the HSC LAE {\tt ALL} catalog as the {\tt unforced} LAEs. The examples of {\tt forced} and {\tt unforced} LAEs are shown in Figure \ref{fig_lae_examples}. As shown in the top-right panels of Figure \ref{fig_lae_examples}, the {\tt unforced} LAEs have a $\simeq0\!\!^{\prime\prime}2-0\!\!^{\prime\prime}3$ spatial offset between centroids in NB and BB. 

\end{enumerate}

In total, we identify $2,230$ and $1,045$ LAE candidates in the HSC LAE {\tt ALL} and {\tt forced} catalogs, respectively. Table \ref{tab_nbdata} presents the numbers of LAE candidates in each field. The machine-readable catalogs of all the LAE candidates will be provided on our project webpage at http://cos.icrr.u-tokyo.ac.jp/rush.html. The photometric properties of example LAE candidates are shown in Table \ref{tab_catalog}. 

As shown in Table \ref{tab_nbdata}, the number of $z\simeq5.7$ LAEs in D-DEEP2-3 appears to be large compared to that of the other $z\simeq5.7$ fields. This may be because the seeing of the ${\it NB}816$ images of D-DEEP2-3 is better than that of the other $z\simeq5.7$ fields. Similarly, the small number of $z\simeq6.6$ LAEs in UD-SXDS may be affected by the seeing size. The number density of LAEs is discussed in the next section. Note that edge regions of UD-COSMOS is overlapped with a flanking field, D-COSMOS (\cite{2017arXiv170405858A}). We find that 30 (7) LAEs in UD-COSMOS are also selected in the HSC LAE {\tt ALL} ({\tt forced}) sample of D-COSMOS. To analyze the D field independently in the following sections, we include the overlapped LAEs in the D-COSMOS sample. 

Figure \ref{fig_nb_color_tile} shows the color-magnitude diagrams for the LAE candidates. The solid curves in the color magnitude diagrams indicate the $3\sigma$ errors of $BB-NB$ color as a function of the NB flux, $f_{\rm NB}$, given by 

\begin{equation}\label{eq_limit}
\pm 3 \sigma_{\rm BB-NB} = -2.5 \log_{10} \big( 1\mp 3\frac{\sqrt{f^2_{\rm 1\sigma NB} + f^2_{\rm 1\sigma BB}}}{f_{\rm NB}} \big), 
\end{equation} 

\noindent where $f_{\rm 1\sigma NB}$ and $f_{\rm 1\sigma BB}$ are the $1\sigma$ flux error in the $z$ and $NB921$ ($i$ and $NB816$) bands for $z\simeq6.6$ ($z\simeq5.7$), respectively. As shown in Figure \ref{fig_nb_color_tile}, the LAE candidates have a significant NB magnitude excess.

\section{Checking the Reliability of Our LAE Selection}\label{sec_check}

Here we check the reliability of our LAE selection. 

\subsection{Spectroscopic Confirmations}

We have conducted optical spectroscopic observations with Subaru/FOCAS and Magellan/LDSS3 for 18 bright LAE candidates with ${\it NB}\lesssim24$ mag. In these observations, we have confirmed 13 LAEs. By investigating our spectroscopic catalog of Magellan/IMACS, we also spectroscopically identify 8 LAEs with ${\it NB}\lesssim24$ mag.  In addition, we find that 75 LAEs are spectroscopically confirmed in literature (\cite{2007ApJS..172..523M, 2008ApJS..176..301O, 2009ApJ...701..915T, 2010ApJ...723..869O, 2012ApJ...760..128M, 2015ApJ...808..139S}; Higuchi et al. in preparation). In total, 96 LAEs have been confirmed in our spectroscopy and previous studies. Using the spectroscopic sample whose number of observed LAEs is known, we estimate the contamination rate to be $\simeq0-30$\%. The details of the spectroscopic observations and contamination rates are given by \citet{shibuya2017b}. 

\subsection{LAE Surface Number Density}

Figure \ref{fig_nb_sd_mf_tile} shows the surface number density (SND) of our LAE candidates and LAEs identified in previous Subaru/SCam NB surveys, SCam LAEs (e.g., \cite{2008ApJS..176..301O,2010ApJ...723..869O}). We find that the SNDs of the {\tt forced} LAEs are comparable to those of SCam LAEs. On the other hand, the SNDs of {\tt unforced} LAEs at $z\simeq6.6$ are higher than that of SCam LAEs. The high SND of the {\tt unforced} LAEs is mainly caused by the color criterion for the HSC LAE {\tt ALL} catalog of $z-{\it NB}921>1.0$ that is less stringent than $z-{\it NB}921>1.8$ (see Section \ref{sec_select}). We also identify SND humps of our {\tt forced} LAEs at $z\simeq6.6$ at the bright-end of ${\it NB}\simeq23$ mag in UD-COSMOS. The presence of such a SND hump has been reported by $z\simeq6.6$ LAE studies (e.g., \cite{2015MNRAS.451..400M}). The significance of the bright-end hump existence in Ly$\alpha$ LFs is $\simeq3\sigma$, which are discussed in \citet{2017arXiv170501222K}. The slight declines in SNDs at a faint NB magnitude of ${\it NB}\gtrsim24.5$ mag would be originated from the incompleteness of the LAE detection and selection. \citet{2017arXiv170501222K} present the SND corrected for the incompleteness. 

Figure \ref{fig_nb_sd_tile} compiles the SNDs of all the HSC UD and D fields. We find that our SNDs show a small field-to-field variation, but typically follow those of the SCam LAEs.

\subsection{Matching Rate of HSC LAEs and SCam LAEs}

The UD-SXDS field has been observed previously by SCam equipped with the ${\it NB}921$ and ${\it NB}816$ filters (\cite{2008ApJS..176..301O,2010ApJ...723..869O}). We compare the catalogs of our selected HSC LAE candidates and SCam LAEs, and calculate the object matching rates as a function of NB magnitudes. The object matching radius is $1^{\prime\prime}$. The object matching rate between the HSC LAEs and SCam LAEs is $\simeq90$\% at a bright NB magnitudes of $\lesssim24$ mag. The high object matching rate indicates that we adequately identify LAEs in our selection processes. However, the matching rate decreases to $\simeq 70$\% at a faint magnitude of $\simeq24.5$ mag. This is due to the shallow depth of the HSC NB fields compared to the SCam ones. \citet{2017arXiv170501222K} discuss the detection completeness of faint LAEs.

\begin{figure*}[t!]
 \begin{center}
  \includegraphics[width=120mm]{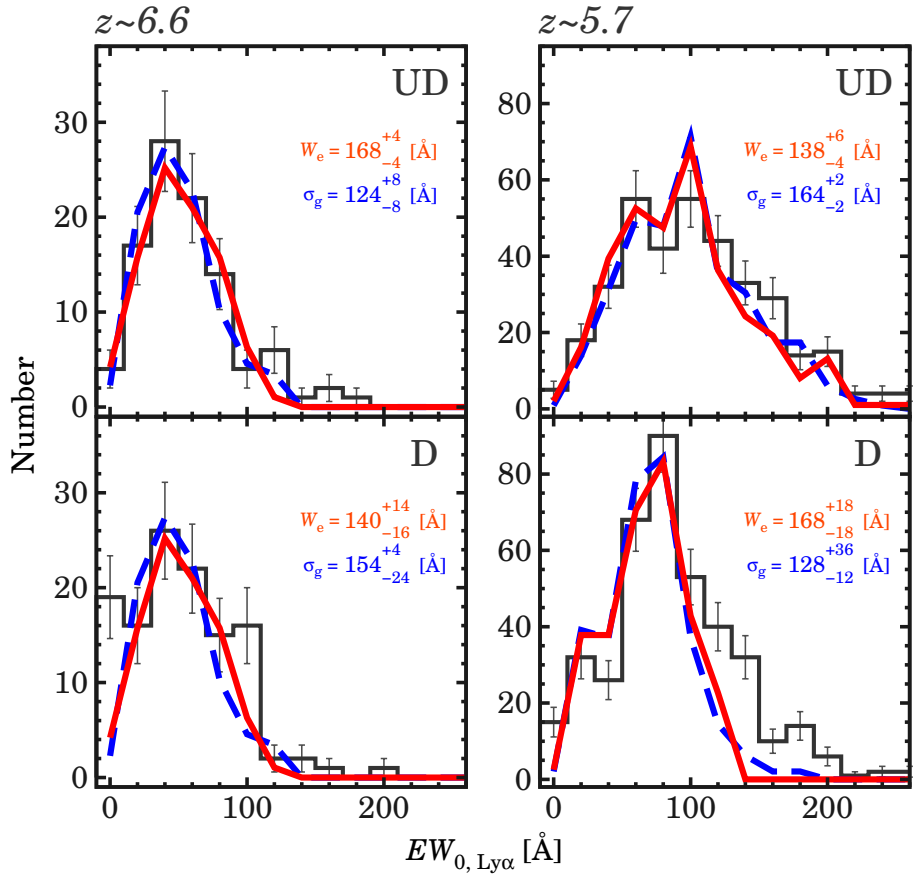}
 \end{center}
 \caption{Ly$\alpha$ EW distribution for the HSC LAEs at $z\simeq6.6$ (left) and $z\simeq5.7$ (right). The top and bottom panels show the UD and D fields, respectively. The thin gray histograms with error bars denote the Ly$\alpha$ EW distributions for the {\tt forced} LAEs. The error bars are given by Poisson statistics from the number of sample LAEs. The red solid and blue dashed lines present the best-fit exponential and Gaussian functions of Equations (\ref{eq_exp}) and (\ref{eq_gauss}), respectively, which are obtained from MC simulations with the $EW_{\rm 0,Ly\alpha}$ uncertainties (see Section \ref{sec_ew_dist} for more details). } \label{fig_hist_ew_tile}
\end{figure*}

\begin{figure*}[t!]
 \begin{center}
  \includegraphics[width=130mm]{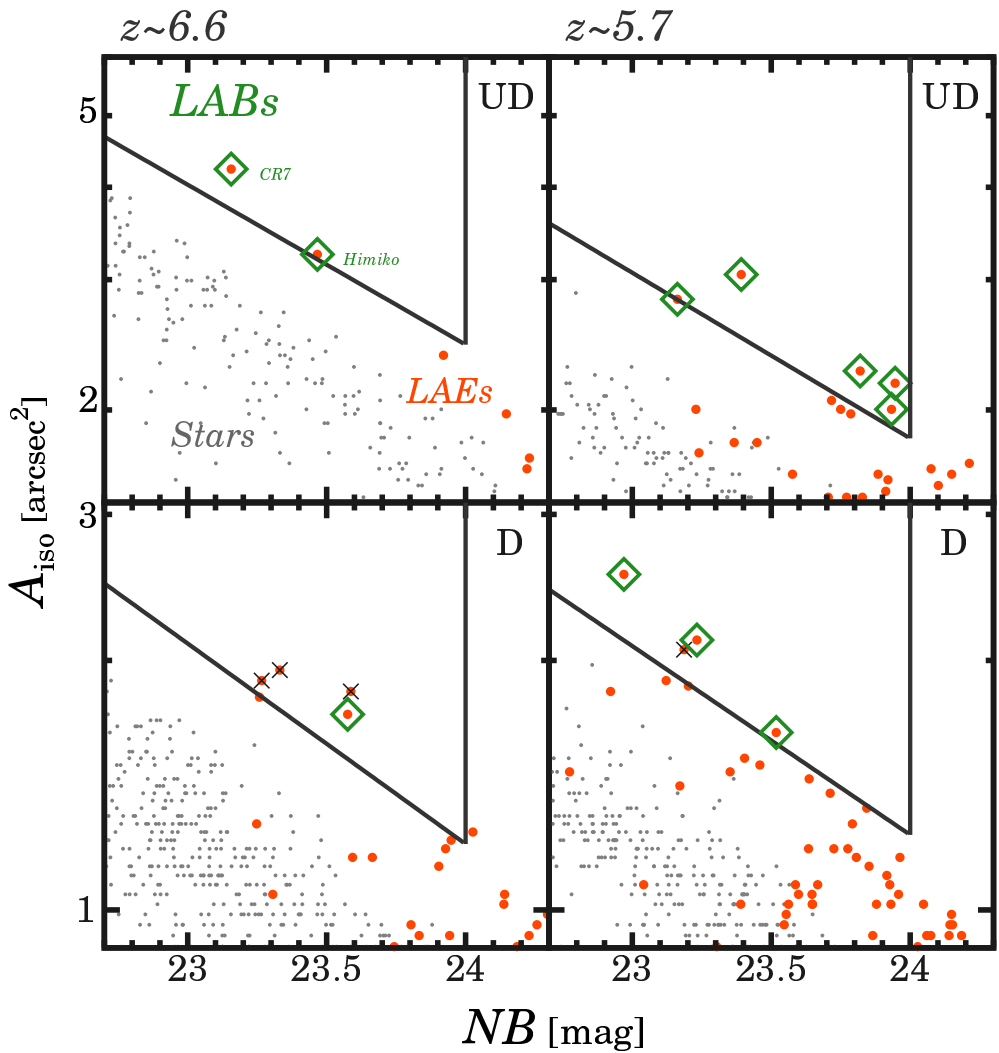}
 \end{center}
 \caption{Isophotal area, $A_{\rm iso}$, as a function of NB magnitude to select LABs at $z\simeq6.6$ (left) and $z\simeq5.7$ (right). The top and bottom panels show the UD and D fields, respectively. The green diamonds denote the LABs. The filled red circles indicate the LAEs in the {\tt forced} catalog. The gray dots represent star-like point sources selected in the HSC NB images. The diagonal and vertical lines denote the LAB selection criteria of $A_{\rm iso}$ and NB magnitude. The diagonal lines are defined by the $2.5\sigma$ deviation from the $A_{\rm iso}$-NB magnitude distribution for the star-like point sources. The filled red circles with a cross indicate unreliable LAB candidates which are affected by e.g., diffuse halos of nearby bright stars. The $z\simeq6.6$ LABs in the UD fields are CR7 (\cite{2015ApJ...808..139S}) and Himiko (\cite{2009ApJ...696.1164O}). }\label{fig_nb_aiso}
\end{figure*}

\begin{longtable}{*{8}{c}}
\caption{Properties of the LABs selected in the HSC NB Data.}\label{tab_lab}
\hline
Object ID & $\alpha$ (J2000) & $\delta$ (J2000) & ${\it NB}_{\rm tot}$ & $UV_{\rm tot}$ & $\log{L_{\rm Ly\alpha}}$ & $EW_{\rm 0, Ly\alpha}$ & $z_{\rm spec}$ \\
 & & &(mag) & (mag) & (erg s$^{-1}$) & (\AA) &  \\
(1) & (2) & (3) & (4) & (5) & (6) & (7) & (8) \\
\hline
\endfirsthead
\endhead
\hline
\endfoot
\hline
\multicolumn{8}{l}{(1) Object ID.} \\
\multicolumn{8}{l}{(2) Right ascension.} \\
\multicolumn{8}{l}{(3) Declination.} \\
\multicolumn{8}{l}{(4) Total magnitudes of ${\it NB}921$- and ${\it NB}816$-bands for $z\simeq6.6$ and $z\simeq5.7$, respectively. } \\
\multicolumn{8}{l}{(5) Total magnitudes of $y$- and $z$-bands for $z\simeq6.6$ and $z\simeq5.7$, respectively.  } \\
\multicolumn{8}{l}{(6) Ly$\alpha$ luminosity. } \\
\multicolumn{8}{l}{(7) Rest-frame equivalent width of Ly$\alpha$ emission line. } \\
\multicolumn{8}{l}{(8) Spectroscopic redshift. } \\
\multicolumn{8}{l}{$^a$ CR7 in \citet{2015ApJ...808..139S}. } \\
\multicolumn{8}{l}{$^b$ Himiko in \citet{2009ApJ...696.1164O}. } \\ 
\multicolumn{8}{l}{$^c$ Spectroscopically confirmed in \citet{shibuya2017b}. } \\
\multicolumn{8}{l}{$^d$ Spectroscopically confirmed in \citet{2012ApJ...760..128M}. } \\
\multicolumn{8}{l}{$^e$ Spectroscopic measurements from the literature. } \\
\endlastfoot
\multicolumn{8}{c}{${\it NB}921$ ($z\simeq6.6$)} \\ \hline
HSC J100058$+$014815$^{a}$ & 10:00:58.00 & $+$01:48:15.14 & 23.25 & 24.48 & 43.9$^{e}$ & $211\pm20$$^{e}$ & 6.604$^{a}$ \\ 
HSC J021757$-$050844$^{b}$ & 02:17:57.58  & $-$05:08:44.64 & 23.50 & 25.40 & 43.4$^{e}$ & $78^{+8}_{-6}$$^{e}$ & 6.595$^{b}$ \\ 
HSC J100334$+$024546$^{c}$ & 10:03:34.66 & $+$02:45:46.56  & 23.61 & 24.97 & 43.5$^{e}$ & $61\pm20$$^{e}$ & 6.575$^{c}$ \\ \hline 
\multicolumn{8}{c}{${\it NB}816$ ($z\simeq5.7$)} \\ \hline
HSC J100129$+$014929 & 10:01:29.07  & $+$01:49:29.81 & 23.47 & 25.87 & 43.4 & $95^{+40}_{-19}$ & 5.707$^{d}$ \\ 
HSC J100109$+$021513 & 10:01:09.72  & $+$02:15:13.45  & 23.13 & 25.77 & 43.6 & $257^{+172}_{-76}$ & 5.712$^{d}$ \\ 
HSC J100123$+$015600 & 10:01:23.84  & $+$01:56:00.46  & 23.94 & 26.43 & 43.3 & $106^{+70}_{-27}$ & 5.726$^{d}$ \\ 
HSC J095946$+$013208 & 09:59:46.73  & $+$01:32:08.45  & 24.16 & 26.12 & 43.1 & $52^{+25}_{-13}$ & --- \\ 
HSC J100139$+$015428 & 10:01:39.94  & $+$01:54:28.34  & 24.11 & 26.58 & 43.2 & $100^{+66}_{-30}$ & --- \\ 
HSC J161927$+$551144 & 16:19:27.73  & $+$55:11:44.70  & 22.88 & 24.86 & 43.7 & $89^{+33}_{-20}$ & --- \\ 
HSC J161403$+$535701 & 16:14:03.82  & $+$53:57:01.25  & 23.53 & 25.32 & 43.4 & $51^{+23}_{-12}$ & --- \\ 
HSC J232924$+$003600 & 23:29:24.85  & $+$00:36:00.34  & 23.62 & 26.48 & 43.4 & $55^{+45}_{-14}$ & --- \\ \hline
\end{longtable}

\begin{figure}[t!]
 \begin{center}
  \includegraphics[width=80mm]{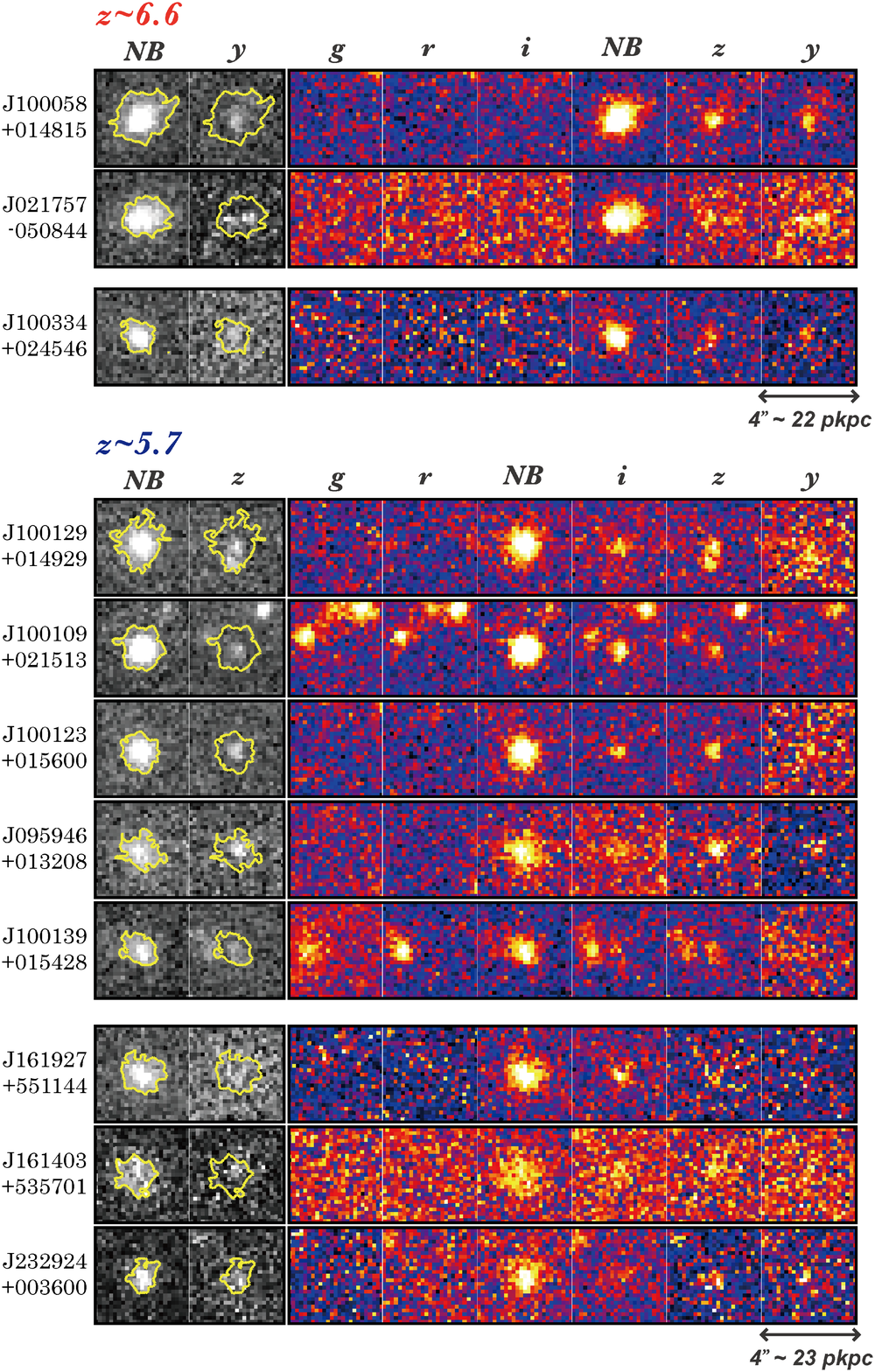}
 \end{center}
 \caption{Postage stamps of the LABs selected with the HSC NB data. The yellow contours indicate isophotal apertures with a threshold of $2\sigma$ sky background noise level. The size of the cutout images is $4^{\prime\prime}\times4^{\prime\prime}$. }\label{fig_lab_postage}
\end{figure}

\section{Results}\label{sec_results}

Here we present the Ly$\alpha$ EW distributions (Section \ref{sec_ew_dist}) and LABs selected with the HSC data (Section \ref{sec_lab_nd}). For the consistency with previous LAE studies, we use the {\tt forced} LAE sample in the following analyses, if not specified. 

\subsection{Ly$\alpha$ EW Distribution}\label{sec_ew_dist}

We present the Ly$\alpha$ EW distributions for LAEs at $z\simeq5.7-6.6$. In a method described in Section \ref{sec_appendix}, we calculate the rest-frame Ly$\alpha$ EW, $EW_{\rm 0, Ly\alpha}$, for the LAEs. The $y$ ($z$) band magnitudes are used for the rest-frame UV continuum emission of $z\simeq6.6$ ($z\simeq5.7$) LAEs. Figure \ref{fig_hist_ew_tile} shows the observed Ly$\alpha$ EW distributions at $z\simeq5.7-6.6$ in the UD and D fields. To quantify these Ly$\alpha$ EW distributions we perform Monte Calro (MC) simulations. The procedure of the MC simulations is similar to that of e.g., \citet{2006PASJ...58..313S}, \citet{2008ApJS..176..301O} and \citet{2014MNRAS.439.1101Z}. First, we generate artificial LAEs in a Ly$\alpha$ luminosity range of $\log{L_{\rm Ly\alpha}}/{\rm erg\ s^{-1}}=42-44$ according to $z\simeq5.7-6.6$ Ly$\alpha$ LFs of \citet{2017arXiv170501222K}. Next, we assign Ly$\alpha$ EW and BB magnitudes to each LAE by assuming that the Ly$\alpha$ EW distributions are the exponential and Gaussian functions (e.g., \cite{2007ApJ...667...79G, 2011ApJ...734..119K,2016ApJ...821L..14O}): 

\begin{equation}\label{eq_exp}
  \frac{{\rm d}N}{{\rm d} EW} = N \exp\Big(-\frac{EW}{W_{\rm e}}\Big), 
\end{equation}

and, 

\begin{equation}\label{eq_gauss}
\frac{{\rm d}N}{{\rm d} EW} = N \frac{1}{\sqrt{2\pi \sigma_{\rm g}^2}} \exp \Big(-\frac{EW^2}{2\sigma_{\rm g}^2} \Big), 
\end{equation}

\noindent where $N$ is the galaxy number, $W_{\rm e}$ and $\sigma_{\rm g}$ are the Ly$\alpha$ EW scale lengths of the exponential and Gaussian functions, respectively. By changing the intrinsic $W_{\rm e}$ and $\sigma_{\rm g}$ values, we make samples of artificial Ly$\alpha$ EW distributions. We then select LAEs based on NB and BB limiting magnitudes and $BB-NB$ colors corresponding to Ly$\alpha$ EW limits which are the same as those of our LAE selection criteria (Section \ref{sec_select}). Finally,  the best-fit Ly$\alpha$ EW scale lengths are obtained by fitting to the artificial Ly$\alpha$ EW distribution to the observed ones.

Figure \ref{fig_hist_ew_tile} presents the Ly$\alpha$ EW distributions obtained in the MC simulations. As shown in Figure \ref{fig_hist_ew_tile}, we find that the Ly$\alpha$ EW distributions are reasonably explained by the exponential and Gaussian profiles. The best-fit scale lengths are summarized in Table \ref{tab_lyaew}. The best-fit exponential (Gaussian) Ly$\alpha$ scale lengths are, on average of the UD and D fields, $153\pm18$\,\AA\ and $154\pm15$\,\AA\ ($146\pm24$\,\AA\, and $139\pm14$\,\AA) at $z\simeq5.7$ and $z\simeq6.6$, respectively. As show in Table \ref{tab_lyaew}, there is no large difference in the Ly$\alpha$ EW scale lengths for the UD and D fields. This no large $EW_{\rm 0,Ly\alpha}$ difference indicates that the results of our best-fit Ly$\alpha$ EW scale lengths does not highly depend on the image depths and the detection incompleteness. In Section \ref{sec_discuss_lyaew}, we discuss the redshift evolution of the Ly$\alpha$ EW scale lengths. 

We investigate LEW LAEs whose intrinsic Ly$\alpha$ EW value, $EW_{\rm 0, Ly\alpha}^{\rm int}$, exceeds $240$\,\AA\ (e.g., \cite{2002ApJ...565L..71M, 2004ApJ...617..707D}). To obtain $EW_{\rm 0, Ly\alpha}^{\rm int}$, we correct for the IGM attenuation for Ly$\alpha$ using the prescriptions of \citet{1995ApJ...441...18M}. In the HSC LAE {\tt ALL} sample, we find that 45 and 230 LAEs have a LEW of $EW_{\rm 0, Ly\alpha}^{\rm int} > 240$\,\AA, for $z\simeq6.6$ and $z\simeq5.7$ LAEs, respectively. These LEW LAEs are candidates of young-metal poor galaxies and AGNs. The fraction of the LEW LAEs in the sample is  $21$\% for $z\simeq5.7$ LAEs. The fraction of LEW LAEs at $z\simeq5.7$ is comparable to that of previous studies on $z\simeq5.7$ LAEs (e.g., $\simeq25$\% at $z\simeq5.7$ in \cite{2008ApJS..176..301O}; $\simeq30-40$\% at $z\simeq5.7$ in \cite{2006PASJ...58..313S}). In contrast, the fraction of LEW LAEs at $z\simeq6.6$ is $4$\% which is lower than that at $z\simeq5.7$. The low fraction at $z\simeq6.6$ might be due to the neutral hydrogen IGM absorbing the Ly$\alpha$ emission. Out of the LEW LAEs, 32 and 150 LAEs at $z\simeq6.6$ and $z\simeq5.7$ exceed $EW_{\rm 0, Ly\alpha}^{\rm int} = 240$ beyond the $1\sigma$ uncertainty of $EW_{\rm 0, Ly\alpha}^{\rm int}$, respectively.

\subsection{LABs at $z\simeq5.7-6.6$}\label{sec_lab_nd}

We search for LABs with spatially-extended Ly$\alpha$ emission. To identify LABs, we measure the NB isophotal areas, $A_{\rm iso}$, for the {\tt forced} LAEs. In this process, we include an {\tt unforced} LAE, Himiko, which is an LAB identified in a previous SCam NB survey (\cite{2009ApJ...696.1164O}). First, we estimate the sky background level of the NB cutout images. Next, we run the {\tt SExtractor} with the sky background level, and obtain the $A_{\rm iso}$ values as pixels with fluxes brighter than the $2\sigma$ sky fluctuation. Note that the NB magnitudes include both fluxes of Ly$\alpha$ and the rest-frame UV continuum emission. Instead of creating Ly$\alpha$ images by subtracting the flux contribution of the rest-frame UV continuum emission, we here simply use the NB images for consistency with previous studies (e.g., \cite{2009ApJ...696.1164O}). 

Using $A_{\rm iso}$ and NB magnitude diagrams, we select LABs which are significantly extended compared to point sources. This selection is similar to that of \citet{2010ApJ...719.1654Y}. Figure \ref{fig_nb_aiso} presents $A_{\rm iso}$ as a function of total $NB$ magnitude. We also plot star-like point sources which are randomly selected in HSC NB fields. The $A_{\rm iso}$ and NB magnitude selection window is defined by a $2.5\sigma$ deviation from the $A_{\rm iso}$-NB magnitude distribution for the star-like point sources. The value of $2.5\sigma$ is applied for fair comparisons with previous studies of e.g., \citet{2009ApJ...693.1579Y} and \citet{2010ApJ...719.1654Y} who have used $\simeq2-4\sigma$. We perform visual inspections for the NB cutout images to remove unreliable LABs which are significantly affected by e.g., diffuse halos of nearby bright stars. 

In total, we identify 11 LABs at $z\simeq5.7-6.6$. Figure \ref{fig_lab_postage} and Table \ref{tab_lab} present multi-band cutout images and properties for the LABs, respectively. As shown in Figure \ref{fig_lab_postage}, these LABs are spatially extended in NB. Our HSC LAB selection confirms that CR7 and Himiko have a spatially extended Ly$\alpha$ emission. Six out of our 11 LABs have been confirmed by our spectroscopic follow-up observations (\cite{shibuya2017b}) and previous studies (\cite{2009ApJ...696.1164O, 2012ApJ...760..128M, 2015ApJ...808..139S}). In Section \ref{sec_discuss_labs}, we discuss the redshift evolution of the LAB number density.

\section{Discussion}\label{sec_discuss}

\begin{table}
  \tbl{Best-fit Ly$\alpha$ EW Scale Lengths}{% 
  \begin{tabular}{ccc}
    \hline
Redshift & $W_{\rm e}$ & $\sigma_{\rm g}$ \\
 & (\AA) & (\AA) \\
(1) & (2) & (3) \\ \hline
6.6 (UD) & $168^{+4}_{-4}$ & $124^{+8}_{-8}$ \\ 
5.7 (UD) & $138^{+6}_{-4}$ & $164^{+2}_{-2}$\\ 
6.6 (D) & $140^{+14}_{-16}$ & $154^{+4}_{-24}$ \\ 
5.7 (D) & $168^{+18}_{-18}$ & $128^{+36}_{-12}$\\ \hline
  \end{tabular}}\label{tab_lyaew}
  \begin{tabnote}
    (1) Redshift of the LAE sample. The parenthesis indicates the UD or D fields. (2) Best-fit Ly$\alpha$ EW scale length of the exponential form. (3) Best-fit Ly$\alpha$ EW scale length of the Gaussian form.  \\
  \end{tabnote}
\end{table}

\subsection{Redshift Evolution of Ly$\alpha$ EW Distribution}\label{sec_discuss_lyaew}

\begin{figure}[t!]
 \begin{center}
  \includegraphics[width=80mm]{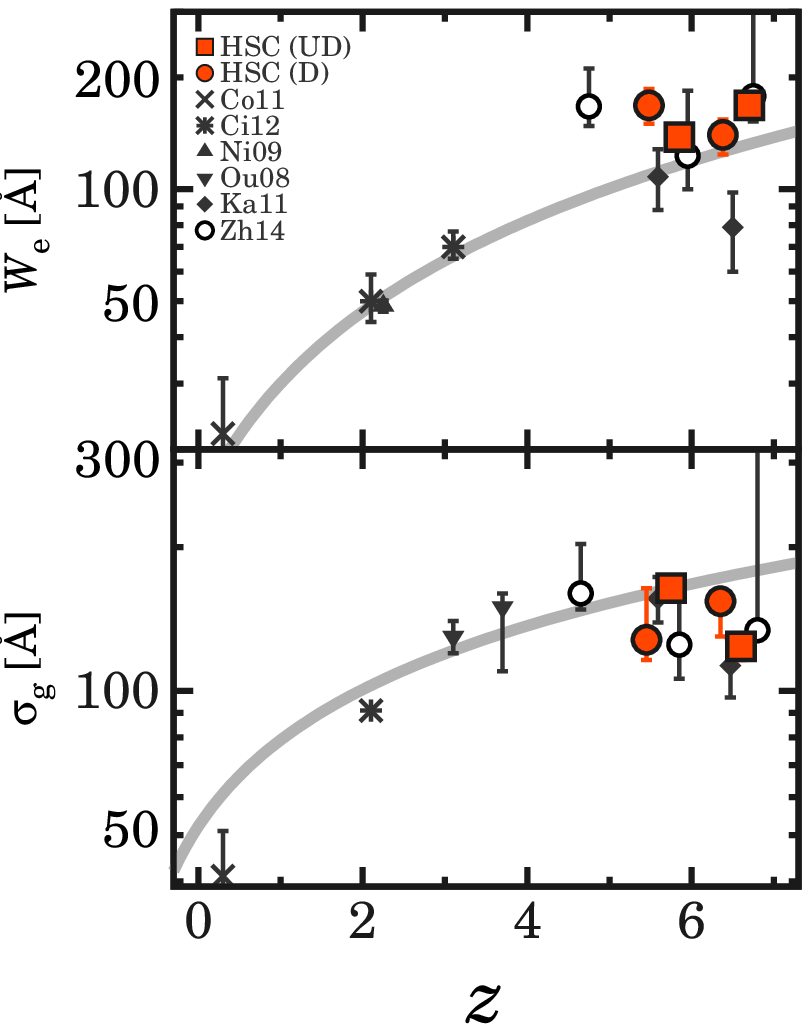}
 \end{center}
 \caption{Redshift evolution of the best-fit Ly$\alpha$ EW scale lengths of the exponential (top) and Gaussian (bottom) functions. The red squares and circles indicate our HSC LAEs in the UD and D fields, respectively. The black symbols are taken from the data points in literature which have been compiled in \citet{2014MNRAS.439.1101Z}  (cross: \cite{2011ApJ...735L..38C}; asterisks: \cite{2012ApJ...744..110C}; filled triangle: \cite{2009A&A...498...13N}; filled inverse triangles: \cite{2008ApJS..176..301O}; filled diamonds: \cite{2011ApJ...734..119K}; open circles: results of Monte-Carlo simulations using data of \cite{2014MNRAS.439.1101Z} and \cite{2010ApJ...725..394H}). The gray curves indicate the best-fit $(1+z)$-form functions obtained in \citet{2014MNRAS.439.1101Z}. }\label{fig_z_ew_scale_tile}
\end{figure}

We discuss the redshift evolution of the Ly$\alpha$ EW scale lengths in a compilation of the results from literature (\cite{2014MNRAS.439.1101Z, 2008ApJS..176..301O, 2009A&A...498...13N, 2010ApJ...725..394H, 2011ApJ...734..119K, 2011ApJ...735L..38C, 2012ApJ...744..110C}). Figure \ref{fig_z_ew_scale_tile} shows the redshift evolution of the Ly$\alpha$ EW scale lengths at $z\simeq0-7$. Our best-fit Ly$\alpha$ scale lengths are comparable to that of \citet{2011ApJ...734..119K} and/or \citet{2014MNRAS.439.1101Z} at $z\simeq5.7-6.6$. The high Ly$\alpha$ EW scale lengths at high-$z$ would indicate that metal-poor and/or less-dusty galaxies with a strong Ly$\alpha$ emission is more abundant at higher-$z$ (e.g., \cite{2011ApJ...728L...2S}). In addition, \citet{2014MNRAS.439.1101Z} have found that the Ly$\alpha$ EW scale length increases towards high-$z$ following a $(1+z)$-form. Our $W_{\rm e}$ and $\sigma_{\rm g}$ values for $z\simeq5.7-6.6$ are also roughly comparable to Zheng et al.'s $(1+z)$-form evolution. However, no significant evolution in the Ly$\alpha$ EW scale lengths from $z\simeq5.7$ to $z\simeq6.6$ is identified in our HSC LAE data, although a possible decline in $\sigma_{\rm g}$ in the UD fields is found. A slight decrease both in $W_{\rm e}$ and $\sigma_{\rm g}$ from $z\simeq5.7$ to $z\simeq6.6$ has been found by \citet{2011ApJ...734..119K}. This sudden decline in the Ly$\alpha$ scale lengths at $z\simeq6.6$ may be caused by the increasing hydrogen neutral fraction in the epoch of the cosmic reionization at $z\gtrsim7$. Note that the Ly$\alpha$ EW scale length measurements would largely depend on BB and NB depths and Ly$\alpha$ EW cuts. Using deeper NB and BB images from the future HSC data release, we will examine the redshift evolution of Ly$\alpha$ scale lengths accurately.

\begin{figure}[t!]
 \begin{center}
  \includegraphics[width=80mm]{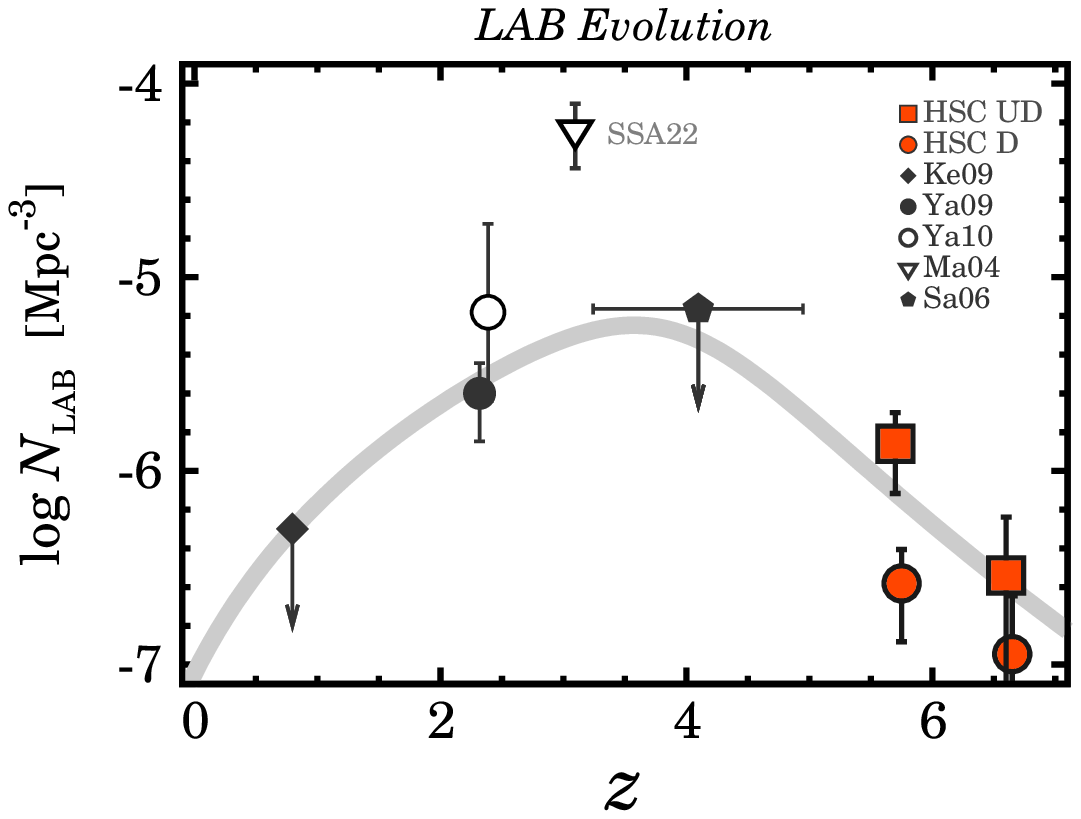}
 \end{center}
 \caption{Redshift evolution of the LAB number density. The filled red squares and filled red circles denote the LABs selected in the HSC UD and D fields, respectively. The error bars are given by Poisson statistics from the LAB number counts. The black symbols show LABs in the literature (filled diamond: \cite{2009AJ....138..986K}; filled circle: \cite{2009ApJ...693.1579Y}; open circle: \cite{2010ApJ...719.1654Y}; filled inverse-triangle: \cite{2004AJ....128..569M}; pentagon: \cite{2006ApJ...648...54S}). All the measurements are based on LABs identified down to the surface brightness limit of $\simeq5\times 10^{-18}$ erg s$^{-1}$ cm$^{-2}$ arcsec$^{-2}$. The gray solid curve represents the best-fit formula of Equation \ref{eq_madau} to the data points expect for the measurement in the  SSA22 proto-cluster region. }\label{fig_z_n_blob}
\end{figure}

\subsection{Redshift Evolution of LAB Number Density}\label{sec_discuss_labs}

We discuss the redshift evolution of the LAB number density, $N_{\rm LAB}$. Figure \ref{fig_z_n_blob} shows $N_{\rm LAB}$ at $z\simeq0-7$ measured by this study and the literature (\cite{2009AJ....138..986K, 2009ApJ...693.1579Y, 2010ApJ...719.1654Y, 2009MNRAS.400L..66M, 2006ApJ...648...54S}). For the plot of the $N_{\rm LAB}$, \citet{2010ApJ...719.1654Y} have compiled $N_{\rm LAB}$ measurements down to an NB surface brightness (SB) limit of $5 \times 10^{-18}$ erg s$^{-1}$ cm$^{-2}$ arcsec$^{-2}$. The SB limits of our HSC NB data are $\simeq5 \times 10^{-18}$ and $\simeq8 \times 10^{-18}$ erg s$^{-1}$ cm$^{-2}$ for the UD and D fields, respectively. Our HSC NB images at least for the UD fields are comparably deep, allowing for fair comparisons with Yang et al.'s $N_{\rm LAB}$ plot. Our $N_{\rm LAB}$ values are $1.4\times10^{-6}$ and $2.9 \times 10^{-7}$ Mpc$^{-3}$ ($2.6\times10^{-7}$ and $1.1\times10^{-7}$ Mpc$^{-3}$) at $z\simeq5.7$ and $z\simeq6.6$ in the UD (D) fields, respectively. The number density at $z\simeq6-7$ is $\simeq 10-100$ times lower than those claimed for LABs at $z\simeq 2-3$ (e.g., \cite{2004AJ....128..569M,2009ApJ...693.1579Y,2010ApJ...719.1654Y}). As shown in Figure \ref{fig_z_n_blob}, there is an evolutional trend that $N_{\rm LAB}$ increases from $z\simeq7$ to $\simeq3$ and subsequently decreases from $z\simeq3$ to $\simeq0$. This trend of the LAB number density evolution is similar to the Madau-Lilly plot of the cosmic SFR density (SFRD) evolution (e.g., \cite{1996MNRAS.283.1388M, 1996ApJ...460L...1L}). Similar to \citet{2016ApJ...821...72S}, we fit the Madau-Lilly plot-type formula, 

\begin{equation}\label{eq_madau}
N_{\rm LAB}(z) = a\times\frac{(1+z)^b}{1 + [(1+z)/c]^d},
\end{equation}
 
\noindent where $a, b, c$, and $d$ are free parameters \citep{2014ARA&A..52..415M} to our $N_{\rm LAB}$ evolution. For the fitting, we exclude \citet{2009MNRAS.400L..66M}'s data point which has been obtained in a overdense region, SSA22. The best-fit parameters are $a=9.1\times10^{-8}$, $b=2.9$, $c=5.0$, and $d=11.7$. 

The similarity of the cosmic SFRD and LAB evolution might indicate that the origin of LABs are related to the star formation activity. As described in Section \ref{sec_intro}, LABs are thought to be formed in physical mechanisms that are connected with the star formation, e.g., the cold gas accretion and the galactic superwinds. The cold gas accretion could  produce the extended Ly$\alpha$ emission powered by the gravitational energy (e.g., \cite{2016MNRAS.457.2318M, 2016ApJ...822...84M,2017ApJ...841...19M}). On the other hand, the superwinds induced by the starbursts in the central galaxies would blow out the surrounding neutral gas, and form extended Ly$\alpha$ nebulae (e.g., \cite{2006Natur.440..644M}). The cold gas accretion rate and the strength of galactic superwinds are predicted to evolve with physical quantities related to the cosmic SFRD (e.g., \cite{2009Natur.457..451D, 2009MNRAS.395..160K}). The comparisons of the cosmic SFRD and LAB evolutions would provide useful hints that LABs are formed in these scenarios. 

However, it should be noted that the LAB selection method is not homogeneous in our comparison of $N_{\rm LAB}$ at $z\simeq0-7$. There is a possibility that the $N_{\rm LAB}$ evolution from $z\simeq7$ to $z\simeq3$ is caused by the cosmological surface brightness dimming effect at high-$z$. The cosmological surface brightness dimming would significantly affect the detection and selection completeness for LABs at high-$z$. To confirm the $N_{\rm LAB}$ evolution and quantitatively compare with the cosmic SFRD, we need to homogenize the selection method for LABs at $z\simeq2-7$ in the future HSC NB data.

\section{Summary and Conclusions}\label{sec_summary}

We develop an unprecedentedly large catalog consisting of LAEs at $z=5.7$ and $6.6$ that are identified by the SILVERRUSH program with the first NB imaging data of the Subaru/HSC survey. The NB imaging data is about an order of magnitude larger than any other surveys for $z\simeq6-7$ LAEs conducted to date. 

Our findings are as follows: 

\begin{itemize} 
\item We identify 2,230 $\gtrsim L^*$ LAEs at $z=5.7$ and $6.6$ on the $13.8$ and $21.2$ deg$^2$ sky, respectively. We confirm that the LAE catalog is reliable on the basis of 96 LAEs whose spectroscopic redshifts are already determined by this program (\cite{shibuya2017b}) and the previous studies (e.g., \cite{2012ApJ...760..128M}). The LAE catalog is presented in this work, and published online. 
\item With the large LAE catalog, we derive the rest-frame Ly$\alpha$ EW distributions of LAEs at $z\simeq5.7$ and $\simeq6.6$ that are reasonably explained by the exponential profile. The best-fit exponential (Gaussian) Ly$\alpha$ scale lengths are, on average of the Ultradeep and Deep fields, $153\pm18$\,\AA\, and $154\pm15$\,\AA\ ($146\pm24$\,\AA\, and $139\pm14$\,\AA) at $z\simeq5.7$ and $z\simeq6.6$, respectively, showing no significant evolution from $z\simeq5.7$ to $z\simeq6.6$. We find 45 and 230 LAEs at $z\simeq6.6$ and $z\simeq5.7$ with a LEW of $EW_{\rm 0,Ly\alpha}^{\rm int}> 240$ \AA\, corrected for the IGM attenuation for Ly$\alpha$. The fraction of the LEW LAEs to all LAEs is $\simeq4$\% and $\simeq21$\% at $z\simeq6.6$ and $z\simeq5.7$, respectively. These LEW LAEs are candidates of young-metal poor galaxies and AGNs. 
\item We search for LABs that are LAEs with spatially extended Ly$\alpha$ emission whose profile is clearly distinguished from those of stellar objects at the $\gtrsim 3\sigma$ level. In the search, we identify 11 LABs in the HSC NB images down to a surface brightness limit of $\simeq5-8 \times 10^{-18}$ erg s$^{-1}$ cm$^{-2}$ which is as deep as data of previous studies. The number density of the LABs at $z\simeq6-7$ is $\sim 10^{-7}-10^{-6}$ Mpc$^{-3}$ that is $\sim 10-100$ times lower than those claimed for LABs at $z\simeq 2-3$, suggestive of disappearing LABs at $z\gtrsim 6$, although the selection methods are different in the low and high-$z$ LABs.
\end{itemize} 

It should be noted that Ly$\alpha$ EW scale length derivation methods and the LAB selections are not homogeneous in a redshift range of $z\simeq0-7$. Using the future $z\simeq2.2, 5.7, 6.6$, and $7.3$ HSC NB data, we will systematically investigate the redshift evolution of Ly$\alpha$ EW scale lengths and $N_{\rm LAB}$ at $z\simeq2-7$ in homogeneous methods.

\section{Appendix: Calculation of Ly$\alpha$ EW}\label{sec_appendix}

In this section, we describe the method to calculate the $EW_{\rm 0,Ly\alpha}$ values. The procedures and the assumption of this method are similar to those of e.g., \citet{2002ApJ...565L..71M}, \citet{2004ApJ...617..707D}, \citet{2007ApJ...667...79G}, \citet{2011ApJ...734..119K}. For the calculation of $EW_{\rm 0,Ly\alpha}$, we assume that LAEs have a $\delta$ function-shaped Ly$\alpha$ line and the flat rest-frame UV continuum emission (i.e. $\beta_\nu = 0$, where $\beta_\nu$ is the UV spectral slope per unit frequency). In such an LAE spectrum, the magnitude, $m$, for a waveband filter with a transmission curve, $T_\nu$, is described as follows:

\begin{equation}\label{eq_ew_mag} 
 48.6 + m = -2.5 \times \log_{10}{ \frac{\int_0^\infty (f_c + f_l \delta(\nu - \nu_\alpha)) T_\nu {\rm d}\nu}{\int_0^\infty T_\nu {\rm d}\nu} }, 
\end{equation} 

\noindent where $f_l$, $f_c$, $\delta(\nu)$, and $\nu_\alpha$ is a Ly$\alpha$ line flux, the flux density of the rest-frame UV continuum emission, the $\delta$ function, and the observed frequency of Ly$\alpha$, respectively. Here we also assume that the Ly$\alpha$ line is located at $9215$\,\AA\ ($8177$\,\AA) which is the central wavelength of the ${\it NB}921$ (${\it NB}816$) filter, for $z\simeq6.6$ ($z\simeq5.7$) LAEs. In this study, we do not take into account the IGM transmission for Ly$\alpha$, if not specified. This is because the IGM transmission for Ly$\alpha$ highly depends on the Ly$\alpha$ line velocity offset from the systemic redshift (e.g., \cite{2013ApJ...765...70H, 2014ApJ...788...74S}). The numerator of the logarithm in Equation (\ref{eq_ew_mag}) corresponds to  

\begin{eqnarray}\label{eq_ew_nume}
& & f_c \int^\infty_{\nu_c} \exp{(-\tau_{\rm eff})} T_\nu {\rm d}\nu + f_c \int_0^{\nu_c} T_\nu {\rm d}\nu + f_l T_\nu (\nu_\alpha) \nonumber \\ 
 &=& f_c B + f_c R + f_l T_\nu (\nu_\alpha).  
\end{eqnarray} 

\noindent In Equation (\ref{eq_ew_nume}), we use $B$, $R$, and $A$ that are defined by equations of 

\begin{eqnarray}\label{eq_ew_bra}
& & B \equiv \int^\infty_{\nu_c} \exp{(-\tau_{\rm eff})} T_\nu {\rm d}\nu, \\ 
& & R \equiv \int_0^{\nu_c} T_\nu {\rm d}\nu, \\ 
& & A \equiv \int_0^\infty T_\nu {\rm d}\nu,  
\end{eqnarray} 

\noindent where $\tau_{\rm eff}$ is the IGM optical depth calculated from analytical models of \citet{1995ApJ...441...18M}. Using Equations (\ref{eq_ew_mag}) and (\ref{eq_ew_nume}), we derive the flux density of the NB and BB filters, $\overline{f_{\rm NB}}$ and $\overline{f_{\rm BB}}$, as follows: 

\begin{eqnarray}\label{eq_ew_fdensity}
 \overline{f_{\rm NB}} &=& 10^{-0.4(m_{\rm NB} + 48.6)} = \frac{f_c (B_{\rm NB} + R_{\rm NB}) + f_l T_{\rm NB} (\nu_\alpha) }{A_{\rm NB}},  \\
  \overline{f_{\rm BB}} &=& 10^{-0.4(m_{\rm BB} + 48.6)} = \frac{f_c (B_{\rm BB} + R_{\rm BB}) }{A_{\rm BB}}. 
\end{eqnarray} 

\noindent The $B$, $R$, and $A$ values with the subscripts of NB (BB) are calculated with the transmission curves of the NB (BB) filters, $T_{\rm NB}$ ($T_{\rm BB}$). In this study, we use magnitudes of the $y$ and $z$ band filters which do not cover the wavelength of Ly$\alpha$ for $z\simeq6.6$ and $z\simeq5.7$ LAEs, respectively, indicating $T_{\rm BB}(\nu_\alpha)=0$. In the case that $m_{\rm BB}$ is fainter than the $1\sigma$ limit, we use the $1\sigma$ limiting magnitude for the $EW_{\rm 0,Ly\alpha}$ calculation. By combining the equations of $\overline{f_{\rm NB}}$ and  $\overline{f_{\rm BB}}$, we obtain $f_c$ and $f_l$,

\begin{eqnarray} 
 f_c &=& \frac{A_{\rm BB} \overline{f_{\rm BB}}}{B_{\rm BB} + R_{\rm BB}} = \overline{f_{\rm BB}}, \\
 f_l &=& \frac{ A_{\rm NB}(B_{\rm BB} + R_{\rm BB})\overline{f_{\rm NB}} - A_{\rm BB}(B_{\rm NB} + R_{\rm NB})\overline{f_{\rm BB}}}{ (B_{\rm BB} + R_{\rm BB})T_{\rm NB} (\nu_\alpha) } \\
 &=& \frac{ A_{\rm NB}\overline{f_{\rm NB}} - (B_{\rm NB} + R_{\rm NB})\overline{f_{\rm BB}}}{ T_{\rm NB} (\nu_\alpha) } \\
 &=& a \times \overline{f_{\rm NB}} - b \times \overline{f_{\rm BB}}. 
\end{eqnarray} 

\noindent Note that $B_{\rm BB} + R_{\rm BB} = A_{\rm BB}$ due to the negligible IGM absorption at the wavelengths of the BB filters. Here we define $a$ and $b$ as

\begin{eqnarray} 
 a &\equiv& \frac{A_{\rm NB}}{T_{\rm NB} (\nu_\alpha)},  \\ 
 b &\equiv& \frac{B_{\rm NB} + R_{\rm NB}}{T_{\rm NB} (\nu_\alpha)}.  
\end{eqnarray} 

\noindent For the HSC ${\it NB}921$ and ${\it NB}816$ filters, the sets of the values are calculated to be $(a,b) \simeq (4.7, 2.3) \times 10^{12}$ and $(a,b) \simeq (5.2, 2.7) \times 10^{12}$, respectively. Using  $f_c$ and $f_l$, we calculate the $EW_{\rm 0,Ly\alpha}$ values via 

\begin{equation}
EW_{\rm 0,Ly\alpha} = \frac{f_l}{f_c} \frac{c}{\nu^2} \frac{1}{1+z}.
\end{equation}
 
To obtain the median values and uncertainties for $EW_{\rm 0,Ly\alpha}$, we perform Monte Carlo (MC) simulations in a method similar to that of e.g., \citet{2006PASJ...58..313S}. In the simulation, we randomly generate a flux density value, $\overline{f_{\rm MC}}$, following a Gaussian probability distribution with an average of $\overline{f}$ and a dispersion of the $1\sigma$ sky background noise, $\overline{f_{\rm 1\sigma}}$, for the NB and BB bands. Here we also randomize $\beta_\nu$ and $\nu_\alpha$ in Gaussian probability distributions with $1\sigma$ dispersions of $\Delta\beta=0.2$ and $\Delta\nu_\alpha={\it FWHM}_{\rm NB}/2.35$, respectively, where ${\it FWHM}_{\rm NB}$ is the FWHM of the NB filters. The dispersion of $\Delta\beta=0.2$ is typical for high-$z$ galaxies (\cite{2014ApJ...793..115B}). In the manner that are the same as described in this section, we calculate a $EW_{\rm 0,Ly\alpha}$ value using $\overline{f_{\rm MC}}$ for NB and BB. In this process, negative values of $f_c$, $f_l$, and $EW_{\rm 0,Ly\alpha}$ are forced to be zero. Such a process is performed 1,000 times for each object. During the iteration, a simulated $EW_{\rm 0,Ly\alpha}$ value is discarded in the case that a ${\it BB}-{\it NB}$ color does not meet the selection criteria of Equations (\ref{eq_6p6select}) and (\ref{eq_5p7select}). Using the set of $EW_{\rm 0,Ly\alpha}$ values obtained from the MC simulations, we calculate the median values and the 16- and 84-percentile errors for $EW_{\rm 0,Ly\alpha}$.

\begin{ack}
We would like to thank James Bosch, Richard S. Ellis, Masao Hayashi, Robert H. Lupton, Michael A. Strauss for useful discussion and comments. We thank the anonymous referee for constructive comments and suggestions. This work is based on observations taken by the Subaru Telescope and the Keck telescope which are operated by the National Observatory of Japan. This work was supported by World Premier International Research Center Initiative (WPI Initiative), MEXT, Japan, KAKENHI (23244025) and (21244013) Grant-in-Aid for Scientific Research (A) through Japan Society for the Promotion of Science (JSPS), and an Advanced Leading Graduate Course for Photon Science grant. The NB816 filter was supported by Ehime University (PI: Y. Taniguchi). The NB921 filter was supported by KAKENHI (23244025) Grant-in-Aid for Scientific Research (A) through the Japan Society for the Promotion of Science (PI: M. Ouchi). NK is supported by the JSPS grant 15H03645. SY is supported by Faculty of Science, Mahidol University, Thailand and the Thailand Research Fund (TRF) through a research grant for new scholar (MRG5980153).

The Hyper Suprime-Cam (HSC) collaboration includes the astronomical communities of Japan and Taiwan, and Princeton University. The HSC instrumentation and software were developed by the National Astronomical Observatory of Japan (NAOJ), the Kavli Institute for the Physics and Mathematics of the Universe (Kavli IPMU), the University of Tokyo, the High Energy Accelerator Research Organization (KEK), the Academia Sinica Institute for Astronomy and Astrophysics in Taiwan (ASIAA), and Princeton University. Funding was contributed by the FIRST program from Japanese Cabinet Office, the Ministry of Education, Culture, Sports, Science and Technology (MEXT), the Japan Society for the Promotion of Science (JSPS), Japan Science and Technology Agency (JST), the Toray Science Foundation, NAOJ, Kavli IPMU, KEK, ASIAA, and Princeton University. 

This paper makes use of software developed for the Large Synoptic Survey Telescope. We thank the LSST Project for making their code available as free software at  http:\/\/dm.lsst.org

The Pan-STARRS1 Surveys (PS1) have been made possible through contributions of the Institute for Astronomy, the University of Hawaii, the Pan-STARRS Project Office, the Max-Planck Society and its participating institutes, the Max Planck Institute for Astronomy, Heidelberg and the Max Planck Institute for Extraterrestrial Physics, Garching, The Johns Hopkins University, Durham University, the University of Edinburgh, Queen's University Belfast, the Harvard-Smithsonian Center for Astrophysics, the Las Cumbres Observatory Global Telescope Network Incorporated, the National Central University of Taiwan, the Space Telescope Science Institute, the National Aeronautics and Space Administration under Grant No. NNX08AR22G issued through the Planetary Science Division of the NASA Science Mission Directorate, the National Science Foundation under Grant No. AST-1238877, the University of Maryland, and Eotvos Lorand University (ELTE) and the Los Alamos National Laboratory.

Based on data collected at the Subaru Telescope and retrieved from the HSC data archive system, which is operated by Subaru Telescope and Astronomy Data Center at National Astronomical Observatory of Japan.
\end{ack}

\bibliographystyle{apj}
\bibliography{reference_prop}

\end{document}